\documentclass[aps,prd,superscriptaddress,nofootinbib,11pt]{revtex4}
\usepackage[english]{babel}
\usepackage[utf8]{inputenc}
\usepackage{graphicx}   
\usepackage{slashed}
\usepackage{epstopdf}
\usepackage{verbatim}   
\usepackage{color}      
\usepackage{subfigure}  
\usepackage{multirow}
\usepackage{hyperref}   
\usepackage{float}
\usepackage{epsfig,rotating}
\usepackage{amsmath,amssymb}
\usepackage{dsfont}
\restylefloat{table}
\numberwithin{equation}{section}

\newcommand{\vx}{\vec{x}}
\newcommand{\vp}{\vec{p}}

\newcommand{\vk}{\vec{k}}

\newcommand{\be}{\begin{equation}}
\newcommand{\ee}{\end{equation}}
\newcommand{\bea}{\begin{eqnarray}}
\newcommand{\eea}{\end{eqnarray}}

\newcommand{\ket}[1]{|#1\rangle}
\newcommand{\bra}[1]{\langle#1|}

\begin{document}
\title{On the origin of  entropy of gravitationally produced  dark matter: \\  the entanglement entropy.}

\author{Mudit Rai}
\email{MUR4@pitt.edu} \affiliation{Department of Physics and
Astronomy, University of Pittsburgh, Pittsburgh, PA 15260}
\author{Daniel Boyanovsky}
\email{boyan@pitt.edu} \affiliation{Department of Physics and
Astronomy, University of Pittsburgh, Pittsburgh, PA 15260}

 \date{\today}

\begin{abstract}
We study the emergence of entropy in gravitational production of dark matter particles, ultra light scalars minimally coupled to gravity and heavier fermions,  from inflation to radiation domination (RD). Initial conditions correspond to dark matter fields in their Bunch-Davies vacua during inflation. The ``out'' states are correlated particle-antiparticle pairs, and the distribution function is found   in both cases. In the adiabatic regime   the density matrix features rapid decoherence by dephasing from interference effects in the basis of ``out'' particle states, effectively reducing it to a diagonal form with a concomitant von Neumann entropy. We show that it is exactly the entanglement entropy obtained by tracing over one member of the correlated pairs. Remarkably, for both statistics  the entanglement entropy is similar to the quantum kinetic entropy in terms of the distribution function  with noteworthy differences stemming from pair correlations.  The entropy and the kinetic fluid form of the energy momentum tensor all originate from decoherence of the density matrix.  For ultra light scalar dark matter, the distribution function peaks at low momentum $\propto 1/k^3$ and the specific entropy  is $\ll 1$. This is a hallmark of a \emph{condensed  phase} but with vanishing field expectation value. For fermionic dark matter the distribution function is nearly thermal and the specific entropy is $\mathcal{O}(1)$ typical of a thermal species. We argue that   the functional form of  the  entanglement entropy is quite general and applies to alternative production mechanisms such as parametric amplification during reheating.
\end{abstract}


\maketitle

\section{Introduction}

 The convergence of evidence for   dark matter (DM) from cosmic microwave background (CMB) anisotropies, galactic rotation curves, gravitational lensing, Bullet cluster,  large scale surveys    and numerical evolution of galaxy formation  is very  compelling.  It is also evident from its properties  that a particle physics  candidate must be sought in extensions beyond the Standard Model (SM). However, a multi decade effort for its direct detection   has not yet led to an unambiguous identification of a (DM) particle\cite{bertone}-\cite{nowimp2}.    A   suitable particle physics candidate must feature a  production mechanism yielding the correct abundance and equation of state, and satisfy the cosmological and astrophysical constraints with a lifetime of at least    the age of the Universe. So far, all of the available evidence is consistent with dark matter interacting solely with gravity.

Among the various   production mechanisms, particle production  as a  consequence of cosmological expansion  is a remarkable phenomenon that has been  studied in pioneering work   in refs.\cite{parker,ford,moste1,birrell,fullbook,parkerbook,mukhabook}. An important aspect of this mechanism is that   if the  particle interacts only  with gravity and no other degrees of freedom, its abundance is determined solely by the particle mass, its coupling to gravity, and the expansion history, independently of hypothetical couplings beyond the (SM). As such, production via cosmological expansion provides a baseline for the abundance and clustering properties of dark matter candidates.

Gravitational production has been studied for various candidates and different cosmological settings: heavy particles produced adiabatically during inflation\cite{heavydm1,heavydm2,heavydm3,kuzmin,kuzmin2,chungfer,ema1,ema2,branreh}, or via inflaton oscillations\cite{vela}, during reheating\cite{hash,vilja1,reheat1,karam,reheat}, or via cosmological expansion during an era with a particular equation of state\cite{vilja2}. More recently the non-adiabatic cosmological  production of ultralight bosonic particles\cite{herring} and heavy fermionic particles \cite{herringfer} were studied during inflation followed by a radiation dominated era.

\vspace{2mm}

\textbf{Motivations, main objectives and brief summary of results.}

Non-adiabatic gravitational production of both   ultra light bosonic dark matter   and a heavier  fermionic dark matter species were studied in references\cite{herring,herringfer} with initial ``in'' conditions during inflation with the respective fields in their Bunch-Davies vacuum state, evolving   to   asymptotic ``out'' particle states in the radiation dominated (RD) era.  The asymptotic ``out'' particle states feature pair correlations and the distribution function is obtained from the Bogoliubov coefficients relating the ``in'' to the ``out'' states which were obtained in these references. Well after the transition from inflation to (RD) and well before matter radiation equality, when the scale factor $a_{eq}\simeq 10^{-4}  \gg a(t) \gg 10^{-17}/\sqrt{m/(\mathrm{eV})}$ there ensues an adiabatic regime during which the Hubble expansion rate $H(t)$ is much smaller than the mass $m$  of the dark matter particle. It is shown in these references that during the adiabatic regime, and after averaging rapid oscillations in interference terms,   the energy momentum tensor of these dark matter particles   feature the kinetic-fluid form. Furthermore, in the case of fermionic dark matter, ref.\cite{herringfer} found that the distribution function features an unexpected near thermality.

These results motivate the main questions addressed in this article: a kinetic-fluid description in terms of a distribution function typically also includes  the entropy\cite{bernstein}, which along with the energy density and pressure provide an effective statistical description of the ``fluid'', as in thermodynamics. In this study we address the   \emph{origin of entropy} associated with this kinetic fluid description.

At \emph{prima facie} the question of entropy within the context of gravitational production seems surprising because the ``in'' state of dark matter is the vacuum state during inflation, therefore the density matrix describes a pure state with vanishing entropy. While this is true, the study in refs.\cite{herring,herringfer} revealed that during the adiabatic regime and in the basis of asymptotic ``out'' particles,  the energy momentum tensor features contributions that evolve on widely different time scales: a slow time scale associated with the cosmological expansion $\simeq 1/H(t)$ and a fast time scale $\simeq 1/m$ associated with the dynamics of the ``out'' particle states. The latter one   is manifest in specific interference terms in pair correlations which \emph{dephase} on the rapid time scale $\simeq 1/m$. As shown explicitly in refs.\cite{herring,herringfer}, the  kinetic-fluid form emerges upon \emph{averaging} these rapidly varying correlations on the longer time scales. The wide separation of these two time scales is precisely the hallmark of the adiabatic regime that sets in well before matter radiation equality. In this article we study whether and how this rapid dephasing phenomena stemming from interference in the asymptotic ``out'' state heralds  a decoherence mechanism, and how such   mechanism entails loss of information and a non-vanishing entropy.

\textbf{Brief summary of results:} Following up on the study of refs.\cite{herring,herringfer}, we consider the non-adiabatic gravitational production of an ultra light complex scalar field minimally coupled to gravity and a heavier fermionic Dirac field under the same set of minimal assumptions considered in these references. The cosmological expansion results in the production of \emph{entangled} correlated asymptotic ``out'' particle-antiparticle pairs of vanishing total momentum.

During  the adiabatic regime, we introduce an effective Schroedinger picture that implements a separation of the widely different time scales, the rapid time scale is included in the time evolution of the density matrix, whereas the slow time scale is associated with  operators. The Bogoliubov transformation that relates the ``in'' to the ``out'' states relates the Schroedinger picture density matrix in the ``in'' basis to the ``out'' basis. Off-diagonal density matrix elements in the ``out'' basis   feature  fast dephasing on   short time scales $\simeq 1/m$,   leading to decoherence and information loss, effectively reducing the density matrix to a diagonal form in this basis, and consequently to  a non-vanishing von Neumann entropy. This rapid dephasing and decoherence in the density matrix   is a direct manifestation of the  interference terms in the energy momentum tensor in the out basis and the emergence of  its  kinetic fluid form.

We show that because gravitational production results in   correlated   particle-antiparticle pairs, the von Neumann entropy resulting from dephasing and decoherence is precisely the \emph{entanglement entropy} obtained by tracing the density matrix over one member of the pairs. Remarkably, the entanglement entropy is similar to the quantum kinetic expression in terms of the distribution function with noteworthy differences arising from the intrinsic pair correlations in the out states.  We find that   the comoving entropy density in terms of the distribution function of produced particles, $N_k$, is given by
\be \mathcal{S} = \pm \frac{1}{2 \pi^2} \int^\infty_0 k^2\,\Big\{(1\pm N_k)\,\ln(1 \pm N_k) \mp  N_k\,\ln N_k   \Big\}  dk \,,\nonumber \ee where $(+)$ is for \emph{real or complex}  bosons and $(-)$ is for each spin/helicity of \emph{Dirac or Majorana} fermions. If the ``out'' states were  \emph{independent} particles and/or antiparticles, complex bosons and Dirac fermions would have twice the number of degrees of freedom of real bosons and Majorana fermions and the entropy would feature an extra factor $2$ when particles are different from antiparticles. The fact that the entropy is the same regardless of whether particles are the same as antiparticles or not is a consequence of the \emph{pair  correlations} of the ``out'' state. These pairs are entangled in momentum (and spin/helicity for fermions), tracing out  any member of the pair yields the same entanglement entropy regardless of whether the member is a particle or an antiparticle. Therefore, the  von Neumann-entanglement- entropy and the kinetic fluid form of the energy momentum are all a  direct consequence of decoherence of the density matrix in the out basis by dephasing.

  We discuss the role of the ``out'' particle basis as a privileged  or ``pointer'' basis,  to describe the   statistical aspects of dark matter, it is preferred by the measurement of the properties of dark matter ``particles''.

For a minimally coupled  ultra light scalar field gravitational production yields a distribution function that is strongly peaked in the infrared\cite{herring}. In this case we find that the specific entropy (entropy per particle) is vanishingly small, this is a hallmark of a \emph{condensed phase} albeit with a vanishing expectation value of the field. For fermionic dark matter, the distribution function is nearly thermal\cite{herringfer} and the specific entropy is $\mathcal{O}(1)$ in agreement with a nearly thermal (but cold) dark matter candidate.

Although we have studied the origin of entropy within these two specific examples, we argue that the emergence of entropy in the production of dark matter from the time evolution of an initial pure state is more generally valid and the mechanism of decoherence by dephasing is common to several alternative proposed  mechanisms of particle production in cosmology.

We note that    cosmological particle production and entanglement entropy have previously been considered for inflationary perturbations\cite{gasp,gasp2,gasperini,prokopec,prokobran,bran1,lello,boyan}, in cosmological particle production\cite{beilok}, and as scenarios of quantum information concepts applied to model cosmologies\cite{martin,ball,mann,machado}. However,  to the best of our knowledge the origin of entropy  has not yet been addressed for non-adiabatic gravitational  production of dark matter during inflation followed by a post inflation radiation dominated cosmology, which is the   focus of our study.

The article is organized as follows:   section (\ref{sec:preli}) summarizes the main assumptions, section (\ref{sec:complex}) studies a complex ultra light scalar dark matter field minimally coupled to gravity, introduces  the method of separation of time scales, obtains the energy momentum tensor and  the density matrix in the out basis, analyzes decoherence by dephasing  and the entanglement entropy. Section (\ref{sec:fdm}) studies fermionic dark matter specifically to understand how particle statistics affects the entanglement entropy.  Section (\ref{sec:discussion}) provides a discussion of various related aspects and arguments for the generality of our results. Section (\ref{sec:conclusions}) summarizes our conclusions and poses new questions. Various appendices supplement technical details.

For self-consistency,  completeness and continuity of presentation,  sections (\ref{sec:complex}) and (\ref{sec:fdm}) include some of the most relevant technical aspects that are discussed in greater detail in refs.(\cite{herring,herringfer}).

\section{Preliminaries:}\label{sec:preli}

We consider a similar cosmological setting as in refs.\cite{herring,herringfer}, namely a spatially flat Friedmann-Robertson-Walker cosmology in conformal time $\eta$ with metric
\be g_{\mu \nu}(\eta) = a^2(\eta)\,\mathrm{diag}(1,-1,-1,-1)\,.  \label{metric} \ee The    assumptions adopted from these references are: \textbf{i:)} the dark matter particle only interacts with gravity but no other degrees of freedom and the dark matter field does not develop an expectation value, \textbf{ii:)} instantaneous transition from inflation to a post-inflation  radiation dominated era, motivated by the consideration of modes that are super-Hubble at the end of inflation, \textbf{iii:)} we take the cosmological dynamics as a \emph{background}: during inflation it is determined by the inflaton field, and during radiation domination (RD) by the more than $\simeq 100$ degrees of freedom of the (SM) (and beyond), \textbf{iv:)} we take all dark matter fields to be in their (Bunch-Davies) vacuum state during inflation.

The inflationary stage is described by a   de Sitter space time (thereby neglecting slow roll corrections)   with a scale factor
\be a(\eta) = -\frac{1}{H_{dS}(\eta-2\eta_R)} \,,\label{adS} \ee where $H_{dS}$ is the Hubble constant during de Sitter and $\eta_R$ is the (conformal) time at which the de Sitter stage transitions to the (RD) stage.

During the     (RD)  stage
\be H(\eta) = \frac{1}{a^2(\eta)}\frac{d a(\eta)}{d\eta} = 1.66 \sqrt{g_{eff}}\,\frac{T^2_0}{M_{Pl}\,a^2(\eta)}\,, \label{hrd}\ee where $g_{eff}$ is the effective number of ultrarelativistic degrees of freedom, which varies in time as different particles become non-relativistic. We take $g_{eff}=2$ corresponding to radiation today. As discussed in references \cite{herring,herringfer}  by taking $g_{eff}=2$ for a fixed dark matter particle mass, one   obtains a \emph{lower bound} on the (DM) abundance and equation of state,  differing  by a factor of $\mathcal{O}(1)$ from the   abundance if the (RD) era is dominated only by (SM) degrees of freedom.  This discrepancy is not relevant for   our study on the origin of entropy.

 With this approximation the scale factor during radiation domination is given by
\be a(\eta)=  H_R\,\eta \,,  \label{Crdmd}\ee with
\be H_R= H_0\,\sqrt{\Omega_R}\simeq 10^{-35}\,\mathrm{eV}   \,,  \label{Hs}\ee and matter radiation equality occurs at
\be a_{eq}= \frac{\Omega_R}{\Omega_M} \simeq 1.66\,\times 10^{-4}  \,.\label{rands}\ee

The result (\ref{Hs}) corresponds to  the value of the fraction density $\Omega_R$ \emph{today}, thereby neglecting the change in the number of degrees of freedom contributing to the radiation density fraction.   For $g_{eff}$ effective ultrarelativistic degrees of freedom, eqn. (\ref{Hs}) must be multiplied by $\sqrt{g_{eff}/2}$. However, as discussed in references\cite{herring,herringfer} accounting for ultrarelativistic degrees of freedom of the (SM) at the time of the transition between inflation and (RD)  modifies the final abundance by a factor of $\mathcal{O}(1)$ and affects the entropy only at a quantitative level by factors of $\mathcal{O}(1)$.

   We require that the scale factor and the Hubble rate be continuous across the transition from inflation to (RD) at  conformal time $\eta_R$, and  assume (self-consistently)  that the transition occurs deep in the (RD) era so that $a(\eta_R) = H_R\,\eta_R \ll a_{eq}$. Continuity of the scale factor and Hubble rate at the instantaneous reheating time results in that the energy density    is continuous at the transition\cite{herring,herringfer}.

 Using   $H(\eta) = a'(\eta)/a^2(\eta)$,  continuity of the scale factor and Hubble rate at $\eta_R$  imply that
\be a_{dS}(\eta_R) = \frac{1}{H_{dS}\,\eta_R}= H_R\,\eta_R ~~;~~ H_{dS} = \frac{1}{H_R\,\eta^2_R} \,, \label{transition} \ee yielding
\be \eta_R = \frac{1}{\sqrt{H_{dS}\,H_R}}\,.  \label{etaR}\ee

Constraints from Planck\cite{planck2018} on the tensor-to-scalar ratio yield the following upper bound on the scale of inflation $H_{dS}$,

\be H_{dS}/M_{Pl} < 2.5\times 10^{-5} ~~(95\%)\,\mathrm{CL} \,. \label{planckcons}\ee We take as a representative value $H_{dS} = 10^{13}\,\mathrm{GeV}$, from which it follows that
\be a_{dS}(\eta_R) =H_R\,\eta_R = \sqrt{\frac{H_R}{H_{dS}}} \simeq 10^{-28}\ll a_{eq} \,, \label{scalefac}\ee consistently with our assumption that the transition from inflation occurs deep in the (RD) era.

With $H_{dS} \simeq 10^{13}\,\mathrm{GeV},H_R \simeq 10^{-35}\,\mathrm{eV}$ it follows that
$\eta_R \simeq 10^{6}/\mathrm{(eV)}$. In our analysis we will consider solely modes that are super-Hubble at the end of inflation, namely with comoving wavevectors $k$  such that
\be k \, \eta_R \ll 1 \,,\label{shubmod}\ee corresponding to comoving wavelengths $\lambda \gg \mathrm{few}\,\mathrm{mts}$. Therefore, all scales of cosmological relevance today correspond to super-Hubble  wavelengths at the end of inflation.

The consideration of solely super-Hubble modes provides an \emph{a priori} justification for the assumption of an instantaneous transition from inflation to  (RD). These modes feature very slow dynamics and in principle are  causally disconnected from microphysical processes, such as collisional thermalization,  occurring on sub-Hubble scales. These considerations suggest that these cosmologically relevant modes are insensitive to the reheating dynamics post-inflation, thereby bypassing the model dependence of reheating mechanisms\cite{reheat1,reheat} and the rather uncertain dynamics of thermalization of (SM) degrees of freedom, which     depends on couplings and   non-equilibrium aspects.

\section{Complex Scalar Fields}\label{sec:complex}

We begin by  considering an ultra light complex scalar field  $\phi$ minimally coupled to gravity, generalizing the study in ref.\cite{herring}. The action in comoving coordinates is given by
\be
S=\int d^{3}xdt\sqrt{-g}\,\Bigg\{\frac{\partial\,\phi^{\dagger}}{\partial t}\,\frac{\partial\,\phi}{\partial t}-\frac{1}{a^2}\,\nabla\phi^{\dagger}\nabla\phi  - m^{2} \,\phi^{\dagger}\,\phi\Bigg\}\,.
\ee
Changing coordinates to conformal time $\eta$ with metric (\ref{metric}),    conformally rescaling the scalar field
\be \phi(\vx,\eta)=\frac{\chi(\vx,\eta)}{a(\eta)}\,,\label{confscalar} \ee  and after discarding a total surface term  the action becomes
\be
S=\int d^{3}x d\eta\,\Bigg\{\chi^{\dagger\,'}\chi'-\nabla\chi^{\dagger}\nabla\chi-M^{2}(\eta)\chi^{\dagger}\chi \Bigg\}
\label{actionconf} \ee where $' \equiv \frac{d}{d\eta} $, and
\be
M^{2}(\eta)=m^{2}a^{2}(\eta)-\frac{a''(\eta)}{a(\eta)} \,. \label{massconf}
\ee

Quantization of the complex scalar field in a comoving volume $V$ is achieved by writing

\begin{equation}
\chi(\vec{x},\eta)=\frac{1}{\sqrt{V}}\,\sum_{\vk}\Big[a_{\vk}\,g_{k}(\eta)\,e^{-i\vk\cdot\vx}+b_{\vk}^{\dagger}\,g_{k}^{*}(\eta)\,e^{i\vk\cdot\vx}\Big]\,,\label{quantchi}
\end{equation}
where the mode functions $g_k(\eta)$ obey the equations of motion

\be g^{''}_k(\eta) + \Big[k^2+m^2 \,a^2(\eta)- \frac{a''(\eta)}{a(\eta)}\,
\Big]g_k(\eta)=0\,,  \label{eomconf}\ee

 and satisfy the   Wronskian
conditions

\be
g^{'}_{k}(\eta)\,g_{k}^{*}(\eta)-g_{k}(\eta)\,{g'}_{k}^{*}(\eta)=-i \,,\label{wronsk}
\ee
which imply canonical commutation relations for the annihilation and creation operators in the expansion (\ref{quantchi}).

\subsection{  ``In-out'' states,  adiabatic mode functions  and particle states. }\label{subsec:asy}

The mode equation   (\ref{eomconf}) can be written in the more familiar form as
 \be -\frac{d^2}{d\eta^2}\, g_k(\eta) + V(\eta)g_k(\eta) = k^2 g_k(\eta) ~~;~~ V(\eta) = -m^2 a^2(\eta) +  \frac{a''(\eta)}{a(\eta)} \,,\label{scheqn}\ee namely a Schroedinger   equation for a wave function $g_k$ with a potential $V(\eta)$ and ``energy'' $k^2$. The potential $V(\eta)$ and/or its derivative are discontinuous at the transition $\eta_R$;  however  $g_k(\eta)$  and $g'_k(\eta)$ are continuous at $\eta_R$. Defining
 \be g_k(\eta) = \Bigg\{ \begin{array}{c}
                   g^<_k (\eta) ~~; ~~ \mathrm{for} ~~;~~ \eta < \eta_R\\
                   g^>_k (\eta) ~~; ~~ \mathrm{for} ~~;~~ \eta > \eta_R\\
                 \end{array} \,,\label{geta} \ee  the matching conditions are
\bea g^<_k(\eta_R) & = & g^>_k(\eta_R) \nonumber \\
\frac{d}{d\eta} g^<_k(\eta)\Big|_{\eta_R} & = &     \frac{d}{d\eta} g^>_k(\eta)\Big|_{\eta_R} \,.    \label{matchcond} \eea

As   discussed in ref.\cite{herring}   these continuity conditions on the mode functions, along with the continuity of the scale factor and Hubble rate     ensure  that the energy density is \emph{continuous} at the transition from inflation to (RD).

 \subsubsection{Inflationary stage:}

 \vspace{1mm}

We consider that the (DM)  scalar field is in the Bunch-Davies vacuum state during the inflationary stage, which corresponds to the mode functions $g_k(\eta)$ fulfilling the boundary condition
\be g_k(\eta) ~~~ _{ \overrightarrow{\eta \rightarrow -\infty}} ~~~ \frac{e^{-ik\eta}}{\sqrt{2k}} \,, \label{BD} \ee and the Bunch-Davies vacuum state $\ket{0_I}$ is such that
\be a_{\vk}\ket{0_I}=0 ~~;~~ b_{\vk}\ket{0_I} =0  \, \, \forall  \vk \,.\label{BDvac} \ee  We refer to this vacuum state as the \emph{in} vacuum.

During the de Sitter stage ($\eta < \eta_R$), with the scale factor given by eqn. (\ref{adS}), the mode equation becomes
\be \frac{d^2}{d\tau^2}\,g^{<}_k(\tau)+\Big[ k^2 - \frac{\nu^2-1/4}{\tau^2}\Big]\,g^{<}_k(\tau) =0 \,, \label{modeqndS}\ee where
\be \tau = \eta-2\eta_R~~;~~ \nu^2 =                                    \frac{9}{4}-\frac{m^2}{H^2_{dS}}
                                     \,. \label{dSparas}\ee The solution with the boundary
condition (\ref{BD})   fulfilling the Wronskian condition (\ref{wronsk})  is given by

\be g^{<}_k(\tau) = \frac{1}{2}\,\sqrt{-\pi \tau}\,\, e^{i\frac{\pi}{2}(\nu+1/2)}\,H^{(1)}_\nu(-k\tau) \label{BDsolution}\ee where $H^{(1)}_\nu$ is a Hankel function. For ultra light dark matter with the correct abundance, the result of ref.\cite{herring} yields $m\simeq 10^{-5}\,\mathrm{(eV)}$, therefore, with $H_{dS}\simeq 10^{13}\,\mathrm{GeV}$ it follows that  $m/H_{dS}   \ll 1 $, hence we can  take $\nu = 3/2$,    yielding
\be g^{<}_k(\tau) =
 \frac{e^{-ik\tau}}{\sqrt{2k}}\,\Big[ 1-\frac{i}{k\tau}\Big] \,.  \label{BDsols2}\ee

  As mentioned in the previous section, we  consider only comoving wavelengths that are \emph{well outside} the Hubble radius at the end of inflation, namely fulfilling the condition (\ref{shubmod}), these describe all  the relevant astrophysical scales today.

  In summary, the  ``\emph{in}''  state is the Bunch-Davies vacuum  defined by equation (\ref{BDvac}) and the mode functions (\ref{BDsols2}) during the   de Sitter inflationary stage.

\subsubsection{Radiation   dominated era:}

During the radiation era  for $\eta > \eta_R$, with $a(\eta) = H_R\eta$    we set $a''=0$, and the mode equation (\ref{eomconf})  becomes
\be  \frac{d^2}{d\eta^2}g^{>}_k(\eta)+\Big[k^2+m^2 \,H^2_R \,\eta^2\Big]g^{>}_k(\eta) =0 \,, \label{paracyl}\ee the general solutions of which are linear combinations of  parabolic cylinder functions\cite{herring,gr,as,nist,bateman,magnus}. As ``out'' boundary conditions, we consider particular solutions that describe asymptotically positive frequency ``particle'' states, their complex conjugate describe antiparticles. This identification relies on a Wentzel-Kramers-Brillouin (WKB)  form of the asymptotic mode functions.

 Let us  consider a particular solution of (\ref{paracyl}) of the WKB form\cite{birrell}
\be f_k(\eta) = \frac{e^{-i\,\int^{\eta}_{\eta_R}\,W_k(\eta')\,d\eta'}}{\sqrt{2\,W_k(\eta)}} \,. \label{WKB}\ee Upon inserting this ansatze in the mode equation (\ref{paracyl}) one finds that $W_k(\eta)$ obeys
\be W^2_k(\eta)= \omega^2_k(\eta)- \frac{1}{2}\bigg[\frac{W^{''}_k(\eta)}{W_k(\eta)} - \frac{3}{2}\,\bigg(\frac{W^{'}_k(\eta)}{W_k(\eta)}\bigg)^2 \bigg]\,,  \label{WKBsol} \ee where
\be \omega^2_k(\eta) = k^2+m^2 \,H^2_R \,\eta^2\,. \label{omegak}\ee

 When $\omega_k(\eta)$ is a slowly-varying function of time the WKB eqn. (\ref{WKBsol}) may be solved in a consistent \emph{adiabatic expansion} in terms of derivatives of $\omega_k(\eta)$ with respect to $\eta$ divided by appropriate powers of the frequency,  namely
\be W^2_k(\eta)= \omega^2_k(\eta) \,\bigg[1 - \frac{1}{2}\,\frac{\omega^{''}_k(\eta)}{\omega^3_k(\eta)}+
\frac{3}{4}\,\bigg( \frac{\omega^{'}_k(\eta)}{\omega^2_k(\eta)}\bigg)^2 +\cdots  \bigg] \,.\label{adexp}\ee  We refer to terms that feature $n$-derivatives of $\omega_k(\eta)$ as of n-th adiabatic order.  During the time interval of rapid variations of the frequencies the concept of particle is ambiguous, but at long time the frequencies evolve slowly   and the concept of particle becomes clear\cite{herring}.

 We want to identify ``particles'' (dark matter ``particles'') near the time of matter radiation equality, so that entering in the matter dominated era when $a(\eta) \simeq a_{eq} \simeq 10^{-4}$, we can extract the  energy momentum tensor  associated with these \emph{ particles}.

  The condition of adiabatic expansion  relies on the ratio
\be \frac{\omega^{'}_k(\eta)}{\omega^{2}_k(\eta)} \ll 1 \,. \label{adiacondi}\ee  An upper bound on this ratio  is obtained in the very long wavelength (superhorizon) limit, taking $\omega_k(\eta) = m \,a(\eta)$,  in a  (RD) cosmology the adiabaticity condition (\ref{adiacondi}) leads to
\be \frac{a'(\eta)}{m \, a^2(\eta)} = \frac{H_R}{m\,a^2(\eta)}  \ll 1  \Longrightarrow a(\eta) \gg \frac{10^{-17}}{\sqrt{m/(eV)}} \,.  \label{alaful}\ee  Therefore,   for  $m \simeq 10^{-5}\,\mathrm{eV}$ corresponding to   $a(\eta) \simeq 10^{-14}$ there is a long period of \emph{non-adiabatic} evolution  since the end of inflation $a(\eta_R) \simeq 10^{-29}  \ll 10^{-14}$, during which the $\omega_k(\eta)$ varies \emph{rapidly}. However, even for an  ultra-light particle with $m  \simeq 10^{-5}\,\mathrm{(eV)}$   the adiabaticity  condition is  fulfilled well before matter-radiation equality.

 The adiabaticity condition (\ref{alaful}) has an important physical interpretation. Since $a'/a^2 = H(t) = 1/d_H(t)$ is the Hubble expansion rate with $d_H$ the Hubble radius (both in comoving time)  it follows that the condition (\ref{alaful}) implies that
 \be  \frac{H(t)}{m} \ll 1 \,\,\, \mathrm{or} \,\,\,  \frac{\lambda_c}{d_H(t)} \ll 1 \,, \label{intadia}\ee  where $\lambda_c$ is the Compton wavelength of the particle. During radiation or matter domination $d_H(t)$ is proportional to the physical particle horizon, therefore the adiabaticity condition is the statement that the Compton wavelength of the particle is much smaller than the physical particle horizon.  The adiabaticity condition becomes less stringent for $k \gg m\,a(\eta)$, in which case it implies that the   comoving de Broglie wavelength is much smaller than the particle horizon. The evolution of the mode functions is non-adiabatic during inflation and for a period  after the transition to (RD)\cite{herring,herringfer}, but becomes adiabatic  well before matter radiation equality.

 During  the adiabatic regime the WKB mode function (\ref{WKB}) asymptotically becomes
\be f_k(\eta) \rightarrow  \frac{e^{-i\,\int^{\eta}\,\omega_k(\eta')\,d\eta'}}{\sqrt{2\,\omega_k(\eta)}}\,. \label{outstates}\ee We refer to the mode functions with this asymptotic boundary condition that fulfill the Wronskian condition

\be  f^{'}_k(\eta)\, f^{*}_k(\eta) - f_k(\eta)\,f^{' *}_k(\eta) = -i \,,\label{wronskf}\ee as ``out''  particle states. As discussed in refs.\cite{herring,herringfer} this criterion is the closest to the particle characterization in Minkowski space-time.

The general solution of equation (\ref{paracyl}) is a linear combination
\be g^>_k(\eta) = A_k\,f_k(\eta) + B_k \,f^*_k(\eta)\,,  \label{generalsol} \ee where $f_k(\eta)$ are the solutions of the mode equation (\ref{paracyl})   with the asymptotic boundary conditions (\ref{outstates}) and $A_k$ and $B_k$ are Bogoliubov coefficients.  Since $g^>_k(\eta)$ obeys the Wronskian condition (\ref{wronsk}) and so does $f_k(\eta)$, it follows that the Bogoliubov coefficients obey
\be |A_k|^2 - |B_k|^2 =1 \,. \label{condiAB}\ee

Using the Wronskian condition (\ref{wronskf}) and the matching condition (\ref{matchcond}),   the Bogoliubov coefficients are determined from the following relations,

\bea A_k & = & i \Big[g^{'\,<}_k(\eta_R)\,f^*_k(\eta_R) - g^{<}_k(\eta_R)\,f^{'\,*}_k(\eta_R)\Big]  \nonumber \\
B_k & = & -i \Big[g^{'\,<}_k(\eta_R)\,f_k(\eta_R) - g^{<}_k(\eta_R)\,f^{'}_k(\eta_R)\Big] \,. \label{ABcoefs}\eea Since the mode functions $g^<_k(\eta)$  also fulfill the Wronskian condition (\ref{wronsk}), it is straightforward to confirm the identity (\ref{condiAB}).

For $\eta > \eta_R$ the field expansion (\ref{quantchi}) yields
\be \chi(\vx,\eta) = \frac{1}{\sqrt{V}}\,\sum_{\vk} \Big[ a_{\vk}\,g^>_k(\eta)\,e^{ i\vk\cdot\vx} + b^\dagger_{\vk}\,g^{*\,>}_k(\eta)\,e^{-i\vk\cdot\vx}\Big] = \frac{1}{\sqrt{V}}\,\sum_{\vk} \Big[ c_{\vk}\,f_k(\eta)\,e^{ i\vk\cdot\vx} + d^\dagger_{\vk}\,f^{*}_k(\eta)\,e^{-i\vk\cdot\vx}\Big] \,, \label{quantchig}\ee where
\be c_{\vk} = a_k \,A_k + b^{\dagger}_{-\vk}\,B^*_k ~~;~~ d^{\dagger}_{\vk} = b^{\dagger}_{\vk}\,A^*_k + a_{-\vk}\,B_k \,. \label{bops} \ee We refer to $c_{\vk},d_{\vk}$ and $c^\dagger_{\vk},d^\dagger_{\vk}$ as the annihilation and creation operators of \emph{out particle and antiparticle} states respectively and the mode functions $f_k(\eta)$ as defining the out basis. These operators  obey canonical quantization conditions as a consequence of the relation (\ref{condiAB}) and are time independent because the mode functions $f_k(\eta)$ are exact solutions of the equations of motion.  The expectation values of bilinears in $c,d$ in the Bunch-Davies vacuum state $|0_I\rangle$   (\ref{BDvac})  are obtained from the relations (\ref{bops}), we find
\be  \bra{0_I}c^\dagger_{\vk}~ c_{\vk'}\ket{0_I} = |B_k|^2\,\delta_{\vk,\vk'} ~~;~~ \bra{0_I}d^\dagger_{\vk} ~ d_{\vk'}\ket{0_I} = |B_k|^2\,\delta_{\vk,\vk'}~~;~~ \bra{0_I}c^\dagger_{\vk} ~ d^\dagger_{-\vk'}\ket{0_I} =  B_k\,A^*_k \,\delta_{\vk,\vk'} \label{parts}\ee with all others vanishing. In particular the number of \emph{out}-particles and anti-particles are  given by
\be  {N}_k = \bra{0_I}c^\dagger_{\vk}  ~ c_{\vk}\ket{0_I} = |B_k|^2= \overline{N}_k = \bra{0_I}d^\dagger_{\vk} ~ d_{\vk}\ket{0_I}\,.  \label{number}\ee We identify $ {N}_k=\overline{N}_k$ with the number of dark matter particles and antiparticles produced \emph{asymptotically} from cosmic expansion. Gravitational production yields the same number of particles as antiparticles. Only in the asymptotic adiabatic regime   can $ {N}_k$ be associated with the number of \emph{particles} (for a more detailed discussion on this point see ref.\cite{herring}).

It remains to obtain the solutions $f_k(\eta)$ of the mode equations (\ref{paracyl}) with asymptotic ``out'' boundary condition (\ref{outstates}) describing asymptotic particle states.

It is convenient to introduce the dimensionless variables
\be x = \sqrt{2mH_R}\, \eta ~~;~~ \alpha = -\frac{k^2}{2mH_R}\,, \label{weberparas} \ee in terms of which the equation (\ref{paracyl}) becomes Weber's equation\cite{as,nist,bateman,magnus}
\be \frac{d^2}{dx^2}\,f(x) +\Big[\frac{x^2}{4}-\alpha \Big]f(x) =0 \,.\label{webereq}\ee
  The solution that satisfies the Wronskian condition (\ref{wronskf}) and features the asymptotic ``out-state'' behavior (\ref{outstates}) with $\omega^2_k(\eta) = \frac{x^2}{4}-\alpha$,  has been obtained in ref.(\cite{herring}) in terms of Weber's function $W[\alpha;x]$\cite{gr,as,nist}. It is  given by

\be f_k(\eta) = \frac{1}{(8mH_R)^{1/4}}\,\Big[\frac{1}{\sqrt{\kappa}}\,W[\alpha;x] -i\sqrt{\kappa}\,W[\alpha;-x]  \Big]~~;~~ \kappa = \sqrt{1+e^{-2\pi|\alpha|}}-e^{-\pi|\alpha|}\,.\label{fconf}\ee

 The Bogoliubov coefficients are obtained from eqns. (\ref{ABcoefs}), where   the mode functions during the de Sitter era, $g^<_k(\eta)$, are given by eqn. (\ref{BDsols2}) (with $\tau = \eta-2\eta_R$). Here we just quote the result for $|B_k|^2$  referring the reader to \cite{herring} for details. In terms of the variable
 \be z = \frac{k}{[2mH_R]^{1/2}}\,, \label{zdef} \ee it is given by
\be {N}_k = |B_k|^2      \simeq  \frac{1}{16\sqrt{2}}~~\bigg( \frac{H_{dS}}{m} \bigg)^2 \frac{D(z)}{z^3} \,. \label{nofkmcfin}\ee where
\be D(z)= \sqrt{1+e^{-2\pi z^2}}~~ \Bigg|\frac{\Gamma\Big(\frac{1}{4}- i\frac{z^2}{2}\Big) }{ \Gamma\Big(\frac{3}{4}- i \frac{z^2}{2}\Big)} \Bigg|\,.  \label{Dofz}\ee This function is analyzed in ref.\cite{herring} but the only properties that are relevant for our discussion are that $D(0) \simeq 4.2$ and that $D(z) \rightarrow \sqrt{2}/z $ for $z \gg 1$.  The infrared enhancement  of  $N_k \propto 1/k^3$ and the prefactor $H_{dS}/m \gg 1$ are both consequences of a minimally coupled light scalar field during inflation\cite{herring}  and results in a distribution function that is strongly peaked with  $N_k \gg 1$     for $z \ll \sqrt{H_{dS}/m}$.

\subsection{Heisenberg vs. adiabatic Schrodinger pictures}\label{subsec:slow}
In the adiabatic regime the mode functions $f_k(\eta)$ with ``out'' boundary conditions can be written as
\be f_k(\eta) = \frac{e^{-i\int^\eta \omega_k(\eta')d\eta'}}{\sqrt{2\omega_k(\eta)}}\,\mathcal{F}_k(\eta) ~~;~~ f'_k(\eta) = -i\omega_k(\eta)\, \frac{e^{-i\int^\eta \omega_k(\eta')d\eta'}}{\sqrt{2\omega_k(\eta)}}\, \mathcal{G}_k(\eta) \,,  \label{fastslow}\ee where
\bea \mathcal{F}_k(\eta)  & =  &  e^{-i\,(\xi^{(1)}(\eta)+\xi^{(2)}(\eta)+\cdots)}\,\,\Big[1 + \mathcal{F}^{(1)}_k(\eta)+ \mathcal{F}^{(2)}_k(\eta)+ \cdots \Big]\,,\label{capF}\\  \mathcal{G}_k(\eta)  & =  &  e^{-i\,(\xi^{(1)}(\eta)+\xi^{(2)}(\eta)+\cdots)}\,\,\Big[1 + \mathcal{G}^{(1)}_k(\eta)+ \mathcal{G}^{(2)}_k(\eta)+ \, \, \cdots \Big]\,. \label{capG}
 \eea The functions $\xi^{(n)}$ are real, and $\xi^{(n)}\,;\,\mathcal{F}^{(n)}_k \,;\,\mathcal{G}^{(n)}$ are of n-th adiabatic order and vanish   in the asymptotic long time limit. During the adiabatic regime $\xi^{(n)}\,;\,\mathcal{F}(\eta)\,;\,\mathcal{G}(\eta)$ are \emph{slowly varying} functions of $\eta$, whereas the phase $e^{-i\int^\eta \omega_k(\eta')d\eta'}$ varies rapidly during a Hubble time. To appreciate  this latter  point more clearly, consider the $k=0$ case for which
 the phase is given in comoving time by $m t \simeq m/H(\eta) = m\,a^2/a' \gg 1$, were the last equality follows from the  adiabaticity condition (\ref{alaful}) during (RD). The important point   is that during the adiabatic regime  there is a wide separation of time scales: the expansion time scale $1/H(t)$ is much longer than the microscopic time scale $1/m$, namely $H(t)/m \ll 1$ which is precisely the adiabaticity condition.

 This important point is at the heart of     decoherence of the density matrix by dephasing discussed below.

 With the slow-fast expansion of the out basis modes (\ref{fastslow}) the expansion of the complex field (\ref{quantchig}) in this basis in the Heisenberg representation is given by
\be \chi(\vx,\eta) =  \,\sum_{\vk} \frac{1}{\sqrt{2\omega_k(\eta)\,V} }\,  \Big[ c_{\vk}\,\mathcal{F}_k(\eta)\,e^{- i\int^\eta_{\eta_i} \omega_k(\eta')d\eta'}\,e^{i\vk\cdot\vx} + d^\dagger_{\vk}\,\mathcal{F}^{*}_k(\eta)\,e^{ i\int^\eta_{\eta_i} \omega_k(\eta')d\eta'}\,e^{-i\vk\cdot\vx}\Big]\,, \label{fastslowchi}\ee where $\eta_i$ is some (arbitrary) early scale but well within  the adiabatic regime. We note that a change of $\eta_i$ may be absorbed into a canonical transformation of $c_{\vk},d_{\vk}$.  Let us introduce the \emph{zeroth order} adiabatic Hamiltonian in the out basis
\be H_0(\eta) = \sum_{\vk} \Big[c^\dagger_{\vk}\,c_{\vk}+ d^\dagger_{\vk}\,d_{\vk} \Big] \,\omega_k(\eta)\,. \label{Had}\ee It follows that
\be [H_0(\eta),c_{\vk}] = -\omega_k(\eta)\,  c_{\vk} ~~;~~ [H_0(\eta),d_{\vk}] = -\omega_k(\eta)\,  d_{\vk}\,.  \label{comcd}\ee

 Although $H_0(\eta)$  depends explicitly on time, it  fulfills
\be [H_0(\eta),H_0(\eta')] = 0  ~~ \forall \eta,\eta' \,. \label{commu}\ee  Therefore, associated with $H_0$ we introduce the unitary time evolution operator
\be U_0(\eta,\eta_i) = e^{-i \int^{\eta}_{\eta_i} H_0(\eta')\, d\eta'}\,,  \label{Uzero}\ee and from the commutation relations (\ref{comcd}) it follows that
\be U^{-1}_0(\eta,\eta_i)~ c_{\vk} ~ U_0(\eta,\eta_i) = c_{\vk}\,e^{-i\int^{\eta}_{\eta_i} \omega_k(\eta') d\eta'} ~~;~~ U^{-1}_0(\eta,\eta_i)~  d_{\vk} ~ U_0(\eta,\eta_i) = d_{\vk}\,e^{-i\int^{\eta}_{\eta_i} \omega_k(\eta') d\eta'} \,.\label{cdofeta}\ee We can now write the Heisenberg picture field operator in the out basis (\ref{fastslowchi})  as
\be \chi(\vx,\eta) = U^{-1}_0(\eta,\eta_i)~ \chi_S(\vx,\eta) ~ U_0(\eta,\eta_i) \,, \label{chiS1}  \ee with the \emph{adiabatic Schroedinger} picture field
\be \chi_S(\vx,\eta) =  \sum_{\vk} \frac{1}{\sqrt{2\omega_k(\eta)\,V}}\,  \Big[ c_{\vk}\,\mathcal{F}_k(\eta)\, e^{i\vk\cdot\vx} + d^\dagger_{\vk}\,\mathcal{F}^{*}_k(\eta) \,e^{-i\vk\cdot\vx}\Big] \,. \label{chiSfin}\ee Similarly with the expansion   (\ref{fastslow}) we find
\be \chi'(\vx,\eta) =  U^{-1}_0(\eta,\eta_i) ~ \Pi_S(\vx,\eta) ~  U_0(\eta,\eta_i)\,, \label{Piofeta}\ee where

\be \Pi_S(\vx,\eta) =      \sum_{\vk}  \frac{-i\,\omega_k(\eta)}{\sqrt{2\omega_k(\eta)\,V}}\,  \Big[ c_{\vk}\,\mathcal{G}_k(\eta)\, e^{i\vk\cdot\vx} - d^\dagger_{\vk}\,\mathcal{G}^{*}_k(\eta) \,e^{-i\vk\cdot\vx}\Big]    \,.  \label{PiSofeta}\ee

  This is the Schroedinger picture version of the adiabatic expansion, $\chi_S(\vx,\eta)\,; \,\Pi_S(\vx,\eta)$ evolve  slowly, on time scales $\simeq 1/H(t)$  in the adiabatic regime, whereas the phases $e^{-i\int^\eta_{\eta_i}\omega_k(\eta')\,d\eta'}$ evolve fast, on time scales $1/m$.

  In the Heisenberg picture operators depend on time but states and the density matrix do not. Consider a Heisenberg picture operator  $\mathcal{O}(\vx,\eta)$ and its expectation value in the Bunch-Davis ``in'' state $\ket{0_I}$,
  \be \bra{0_I}\mathcal{O}(\vx,\eta)\ket{0_I} = \bra{0_I} U^{-1}_0(\eta,\eta_i) ~ \mathcal{O}_S(\vx,\eta) ~  U_0(\eta,\eta_i)\ket{0_I} \equiv \mathrm{Tr} \Big[\rho_S(\eta)~\mathcal{O}_S(\vx,\eta) \Big]\,, \label{exval}\ee where we have introduced the adiabatic  Schroedinger picture density matrix
  \be \rho_S(\eta) = U_0(\eta,\eta_i)\ket{0_I} \bra{0_I} U^{-1}_0(\eta,\eta_i)\,.  \label{densmat}\ee Obviously this density matrix describes a pure state since $\rho^2_S(\eta) = \rho_S(\eta)$. This adiabatic Schroedinger picture  effectively separates the fast time evolution, now encoded in the density matrix, from the slow time evolution of the field operators $ \mathcal{O}_S(\vx,\eta) $.

  In Minkowski space time the Schroedinger picture operators $ \mathcal{O}_S(\vx,\eta) $ do not evolve in time whereas the states and the density matrix evolves in time with the usual time evolution operator $e^{-i H t}$. During the adiabatic regime in (RD) cosmology the equivalent Schroedinger picture operators feature a slow residual adiabatic time evolution on the time scales of cosmological expansion.

\subsection{Energy Momentum Tensor}\label{subsec:scalaremt}

For a minimally coupled complex scalar field, the energy momentum tensor is given by

\be
T_{\mu\nu}=\partial_{\mu}\phi^{\dagger}\partial_{\nu}\phi+\partial_{\nu}\phi^{\dagger}\partial_{\mu}\phi
-g_{\mu\nu}\big[g^{\alpha\beta}\partial_{\alpha}\phi^{\dagger}\partial_{\beta}\phi-m^{2}|\phi|^{2}\big]\,.
\label{EMT}
\ee

In conformal time and after the conformal rescaling of the field (\ref{confscalar}) we find (  space-time arguments are implicit)
\be
T_{0}^{0}  =\frac{1}{a^{4}}\Bigg[(\chi'-\frac{a'}{a}\chi)^{\dagger}(\chi'-\frac{a'}{a}\chi)+\nabla\chi^{\dagger}\cdot
\nabla\chi+m^{2}a^{2}|\chi|^{2}\Bigg]\,,\label{tzerozero}
\ee along with

\be
T_{\mu}^{\mu}  = \frac{2}{a^{4}}\Bigg[2m^{2}a^{2}|\chi|^{2}-(\chi'-\frac{a'}{a}\chi)^{\dagger}(\chi'-\frac{a'}{a}\chi)+
\nabla\chi^{\dagger}\cdot\nabla\chi\Bigg]\,.\label{Tmumu} \ee
The Bunch-Davies ``in'' vacuum state is homogeneous and isotropic
therefore the expectation value of the energy momentum tensor in this state features the ideal fluid form
$\bra{0_I}T^{\mu}_{\nu}\ket{0_I} = \mathrm{diag}\big(\overline{\rho}(\eta),-\overline{P}(\eta),-\overline{P}(\eta),-\overline{P}(\eta)\big)$. It proves convenient to extract the homogeneous and isotropic components of the energy momentum tensor as an operator, this is achieved by its averaging over the comoving volume $V$, namely
\be \frac{1}{V} \int  d^3 x \, T^{0}_{0}(\vx,\eta)  = \widehat{\overline{\rho}}(\eta)~~;~~ \frac{1}{V} \int  d^3 x  \, T^{\mu}_{\mu}(\vx,\eta) = \widehat{\overline{\rho}}(\eta)-3\, \widehat{\overline{P}}(\eta)\,,\label{volave}\ee where the hat refers to the operator. Since we are interested in the energy momentum tensor near matter radiation equality well within the adiabatic regime, we obtain these volume averages by implementing two steps: \textbf{i:)} the field $\chi$ is written in the ``out'' basis, namely in terms of the mode functions $f_k(\eta)$  as in eqn. (\ref{quantchig}), \textbf{ii:)} these mode functions are written by separating the slow and fast parts as in eqns. (\ref{fastslow},\ref{fastslowchi}), we find

\begin{align}
 \begin{split}
 \widehat{\overline{\rho}}(\eta)    =  & \frac{1}{2\,V\,a^4(\eta)}\sum_{\vk} \Bigg\{ \Bigg[1+c^\dagger_{\vk}\,c_{\vk}+d^\dagger_{\vk}\,d_{\vk} \Bigg]  \,\Bigg[\Big(|\mathcal{F}|^2+|\mathcal{G}|^2 \Big) \, \omega_k(\eta)   -  i\Big( \frac{a'}{a}\Big)\, \Big(\mathcal{G}^*\,\mathcal{F}-\mathcal{G}\,\mathcal{F}^*\Big) + \Big( \frac{a'}{a}\Big)^2\,\frac{|\mathcal{F}|^2}{\omega_k(\eta)} \Bigg]+ \\& c^\dagger_{\vk}\,d^\dagger_{-\vk}\,e^{2i\int^\eta_{\eta_i} \omega_k(\eta')\,d\eta'}\,  \Bigg[\omega_k(\eta) \,\Big({\mathcal{F}^{*}}^2 - {\mathcal{G}^{*}}^2 \Big)-2i\Big( \frac{a'}{a}\Big) \Big(\mathcal{F}\,\mathcal{G} \Big)^* +\Big( \frac{a'}{a}\Big)^2\,\frac{ {\mathcal{F}^{*}}^2}{\omega_k(\eta)}   \Bigg] + \\&
 c_{\vk}\,d_{-\vk}\,e^{-2i\int^\eta_{\eta_i} \omega_k(\eta')\,d\eta'}\,  \Bigg[\omega_k(\eta) \,\Big({\mathcal{F}}^2 - {\mathcal{G}}^2 \Big)+2i\Big( \frac{a'}{a}\Big) \Big(\mathcal{F}\,\mathcal{G} \Big) +\Big( \frac{a'}{a}\Big)^2\,\frac{ {\mathcal{F}}^2}{\omega_k(\eta)}   \Bigg]    \Bigg\} \label{rhoop}\,,
  \end{split}
 \end{align}
and

\begin{align}
\begin{split}
 \widehat{\overline{\rho}}(\eta)-3\, \widehat{\overline{P}}(\eta) =&\frac{1}{V \,a^{4}(\eta)}\,\sum_{\vk}\Bigg\{ \Big(1+c^\dagger_{\vk}c_{\vk}+d^\dagger_{\vk}d_{\vk}\Big)\,
 \Bigg[ \frac{m^{2}a^2(\eta)}{\omega_k(\eta)} \,|\mathcal{F}|^{2}
\\&+ \omega_k(\eta)\, \big(|\mathcal{F}|^{2}-|\mathcal{G}|^{2}\big)+
   i\Big( \frac{a'}{a}\Big)\, \Big(\mathcal{G}^*\,\mathcal{F}-\mathcal{G}\,\mathcal{F}^*\Big) - \Big( \frac{a'}{a}\Big)^2\,\frac{|\mathcal{F}|^2}{\omega_k(\eta)} \Bigg] \\&+ c^\dagger_{\vk}\,d^\dagger_{-\vk}\,e^{2i\int^\eta_{\eta_i} \omega_k(\eta')\,d\eta'}\,
\Bigg[\frac{{\mathcal{F}^*}^2}{\omega_k}\,\big(m^{2}a^{2}+\omega_{k}^{2}\big)-\frac{1}{\omega_k}\,
\Big(i\omega\,\mathcal{G}^{*}-\frac{a'}{a}\mathcal{F}^{*}\Big)^{2}\Bigg] \\&
+ c_{\vk}\,d_{-\vk}\,e^{-2i\int^\eta_{\eta_i} \omega_k(\eta')\,d\eta'}\,
\Bigg[\frac{{\mathcal{F}}^2}{\omega_k}\,\big(m^{2}a^{2}+\omega_{k}^{2}\big)-\frac{1}{\omega_k}\,
\Big(-i\omega\,\mathcal{G}-\frac{a'}{a}\mathcal{F}\Big)^{2}\Bigg]
\Bigg\}\,.\label{tmumuope}
\end{split}
\end{align}

  The expectation values of these operators in the ``in'' vacuum state are readily obtained from equations (\ref{parts}).

 These expressions   show explicitly that the contributions  that  are    diagonal in the ``out'' basis, namely, $c^\dagger c~;~ d^\dagger d$ are slowly varying, whereas the off-diagonal terms $c\,d~;,c^\dagger d^\dagger$ exhibit the fast varying phases. These rapidly varying  terms are a consequence of the  interference between particle and antiparticle ``out'' states, similar to the phenomenon of \emph{zitterbewegung},  and average out over time scales $\gtrsim 1/m$ leaving only the diagonal contributions to the energy density and pressure\cite{herring}. The energy momentum tensor, as an operator, can also be written  passing to the adiabatic Schroedinger picture as
 \be T^{\mu \nu}(\vx, \eta) = U^{-1}_0(\eta,\eta_i)\,T^{\mu \nu}_{S}(\vx, \eta)\,  U_0(\eta,\eta_i)\,, \label{tmunusch}\ee where $U_0(\eta,\eta_i)$ is the time evolution operator (\ref{Uzero}) removing the fast varying phases in  (\ref{rhoop},\ref{tmumuope}), and $T^{\mu \nu}_{S}(\vx, \eta)$ is the adiabatic Schroedinger picture operator with slow time evolution in the adiabatic regime. In terms of the adiabatic Schroedinger picture density matrix (\ref{densmat}), it follows that
 \be \bra{0_I}T^{\mu \nu}(\vx,\eta)\ket{0_I}= \mathrm{Tr}\Big[ \rho_S(\eta)\,T^{\mu \nu}_{S}(\vx, \eta)\Big] \,.\label{exvalS}\ee
 The rapidly varying phases in the particle-antiparticle interference  terms in the ``out'' basis in (\ref{rhoop},\ref{tmumuope}) suggest that the off diagonal elements of the density matrix $\rho_S(\eta)$ in the ``out'' basis will also feature these rapidly varying phases from particle-antiparticle interference, which average out on time scales $\gtrsim 1/m$. This averaging suggests a process of \emph{decoherence by dephasing}, which is analyzed in detail in the next section.

\subsection{Decoherence of the density matrix: von Neumann and entanglement entropy}\label{subsec:entropy}
In appendix (\ref{app:bogobose}) we show that the `in'' Bunch-Davies vacuum state can be written in terms of the Fock states of the ``out'' basis as (see appendix (\ref{app:bogobose}) for definitions)
\be \ket{0_I} = {\Pi}_{\vk} \sum_{n_{\vk}=0}^\infty \mathcal{C}_{n_{\vk}}(k) \, \ket{n_{\vk};\overline{n}_{\vk}} ~~;~~ \mathcal{C}_{n_{\vk}}(k) = \frac{\Bigg(e^{2i\varphi_-(k)}\,\tanh(\theta_k) \Bigg)^{n_{\vk}}  }{\cosh(\theta_k)} \,, \label{invacout}\ee with
\be |B_k|^2 = \sinh^2(\theta_k)=N_k~~;~~|A_k|^2 = \cosh^2(\theta_k) ~~;~~ \tanh^2(\theta_k) = \frac{N_k}{1+N_k}\,, \label{idenN} \ee  and
\be e^{2i\varphi_-(k)}\,\tanh(\theta_k) = \frac{B^*_k}{A^*_k}\,, \label{bovera}\ee
and the correlated Fock pair states
\be  \ket{n_{\vk};\overline{n}_{-\vk}}  = \frac{\Big(c^{\dagger}_{\vk}\Big)^{n_{\vk}}}{\sqrt{n_{\vk}\, !}}~
\frac{\Big(d^{\dagger}_{-\vk}\Big)^{n_{\vk}}}{\sqrt{n_{\vk}\, !}} \, \ket{0_O} ~~;~~ n_{\vk} = 0,1,2\cdots \,, \label{nouts1}\ee where the ``out'' vacuum state $\ket{0_O}$ is such that
\be c_{\vk}\,\ket{0_O} = d_{\vk}\,\ket{0_O} =0\,.  \label{vaouout1}\ee
We note that the Fock pair states (\ref{nouts1}) are eigenstates of the \emph{pair number operator}
\be \widehat{\mathcal{N}}_{\vk} = \sum_{m_{\vk}=0}^\infty m_{\vk}\,\, \ket{m_{\vk};\overline{m}_{-\vk}}\bra{m_{\vk};\overline{m}_{-\vk}}  \,, \label{pairop} \ee with
\be \widehat{\mathcal{N}}_{\vk} \, \ket{n_{\vk};\overline{n}_{-\vk}} = n_{\vk} \,\ket{n_{\vk};\overline{n}_{-\vk}}~~;~~ n_{\vk} = 0,1,2\cdots \,.  \label{eigenNk}\ee

In this ``out'' basis and in the adiabatic regime prior to matter-radiation equality, the density matrix in the Schroedinger picture (\ref{densmat}) becomes
\be \rho_S(\eta) = \Pi_{\vk} \Pi_{\vp}  \sum_{n_{\vk}=0}^\infty  \sum_{m_{\vp}=0}^\infty \mathcal{C}^*_{m_{\vp}}(p)~ \mathcal{C}_{n_{\vk}}(k)~\ket{n_{\vk};\overline{n}_{-\vk}}\bra{m_{\vp};\overline{m}_{-\vp}}~e^{2i \int^\eta_{\eta_i} \Big[m_{\vp}\,\omega_p(\eta')-n_{\vk}\,\omega_k(\eta') \Big]\,d\eta'} \,. \label{rhosout}\ee The diagonal density matrix elements both in momentum and number of particles, namely  $\vk= \vp~;~ m_{\vp} = n_{\vk}$ are time independent, these describe the ``populations'', whereas the off-diagonal elements describe the coherences. These latter matrix elements vary rapidly in time and average out over time scales $\gg 1/m$. To see this aspect more clearly, and  recognizing that
\be \int^\eta \omega_k(\eta') d\eta' = \int^t E_k(t') dt' ~~ ; ~~ E_k(t) = \sqrt{\frac{k^2}{a^2(t)}+m^2} \label{integt}\ee let us consider  the average
\be \frac{1}{(t_f-t_i)} \,\int^{t_{f}}_{t_i} e^{2i \int^t \Big[m_{\vp}\,E_p(t')-n_{\vk}\,E_k(t') \Big]\,dt'} \,dt ~~;~~ m(t_f-t_i) \gg 1\,.\label{avetime}\ee For example for $\vp = \vk =0$ and $m(t_f-t_i)\gg 1$ the integral yields $\delta_{m_{\vec{0}},n_{\vec{0}}}$. Taking the interval $t_f-t_i$  of the order of the Hubble time $\simeq 1/H(t)$, in the adiabatic regime with $H(t)/ m \ll 1$  the integral yields  $\simeq H/m \ll 1$ for $m_{\vec{0}}\neq n_{\vec{0}}$ and $\mathcal{O}(1)$ for $m_{\vec{0}}= n_{\vec{0}}$. Therefore, the rapidly varying phases effectively average out the coherences over time scales $\simeq 1/m \ll 1/H(t)$  projecting the density matrix to the diagonal elements in the ``out'' basis.

In summary: the rapid dephasing of the off-diagonal matrix elements in the out basis in the adiabatic regime average these contributions on time scales of order $1/m$ which are much shorter than the expansion time scale (Hubble scale) in the adiabatic regime. The rapid dephasing leads to \emph{decoherence} in the ``out'' basis, the time averaging is tantamount to a coarse graining over short time scales leaving effectively a diagonal density matrix   in this basis,  describing a \emph{mixed state} that evolves slowly on the long time scale,
\be \rho^{(d)}_S =  \Pi_{\vk} \big[1-\tanh^2(\theta_k) \big]  \sum_{n_{\vk}=0}^\infty \Big(\tanh^2(\theta_k)\Big)^{n_{\vk}}
\ket{n_{\vk};\overline{n}_{-\vk}}\bra{n_{\vk};\overline{n}_{-\vk}}\,. \label{diagrho}\ee This density matrix is diagonal in the Fock ``out'' basis of correlated --entangled-- particle-antiparticle pairs, and in $\vk$ space, with the diagonal matrix elements representing the probabilities. We note that $\mathrm{Tr}\,\rho^{(d)}_S =1$. The entropy associated with this mixed state can be calculated simply by establishing contact between the density matrix $\rho^{(d)}_S$ and that of  quantum statistical mechanics in equilibrium described by a fiducial Hamiltonian
\be \widehat{\mathcal{H}} = \sum_{\vk} \mathcal{E}_k\, \widehat{\mathcal{N}}_{\vk} \,,\label{fiduH} \ee with $\widehat{\mathcal{N}}_{\vk}$ the pair number operator (\ref{pairop}) with eigenvalues $n_{\vk} = 0,1,2\cdots$,   and the fiducial energy
\be \mathcal{E}_k = - \ln\big[\tanh^2(\theta_k) \big]\,.  \label{fiduenergy}\ee This fiducial Hamiltonian is diagonal in the correlated basis of particle-antiparticle pairs, therefore we identify
\be \rho^{(d)}_S = \frac{e^{-\widehat{\mathcal{H}}}}{\mathcal{Z}}~~;~~ \mathcal{Z}= \mathrm{Tr}\,e^{-\widehat{\mathcal{H}}} \equiv e^{-\mathbb{F}} \,,\label{rhofidu}\ee with $\mathbb{F}$ the fiducial free energy, and
 \be \mathcal{Z} =  \Pi_{\vk} \mathcal{Z}_{\vk}~~;~~ \mathcal{Z}_{\vk} = \frac{1}{\Big[ 1- e^{-\mathcal{E}_{k}}\Big]} = \frac{1}{\Big[ 1- \tanh^2(\theta_k)\Big]} \,.   \label{fiduzexa}\ee
  Obviously the matrix elements of (\ref{rhofidu})  in the pair basis are identical to those of (\ref{diagrho}).

 The von Neumann entropy associated with this mixed state is
 \be S^{(d)} = - \mathrm{Tr}\,\rho^{(d)}_S\,\ln \rho^{(d)}_S \,. \label{vNSd} \ee

 Since $\widehat{\mathcal{H}}$ is diagonal in the basis of the pair Fock states (\ref{nouts1}), so is $\rho^{(d)}_S$. The eigenvalues of $\rho^{(d)}_S$ are the probability for each state of $n_{\vk}$ pairs of momenta $(\vk;-\vk)$, namely
 \be  P_{\vk;n_{\vk}} = \frac{e^{-\mathcal{E}_k\,n_{\vk}}}{\mathcal{Z}_{\vk}}~~;~~\sum_{n_{\vk}=0}^\infty P_{\vk;n_{\vk}}=1\,,\label{probascalar}\ee  therefore  the von Neumann entropy is given by
\be S^{(d)}  = -\sum_{\vk}\sum_{n_{\vk}=0}^\infty P_{\vk;n_{\vk}}\,\ln P_{\vk;n_{\vk}}\,. \label{Sden} \ee

This is equivalent to a simple quantum statistical mechanics problem.   The relation
\be \mathbb{F} = - \ln\mathcal{Z} = U - S^{(d)} ~~;~~ U = \mathrm{Tr} \rho^{(d)}_S \, \widehat{\mathcal{H}}\,, \label{thermorel}\ee   is a direct consequence of the expression (\ref{Sden}) for $S^{(d)}$ and the normalized probabilities $ P_{\vk;n_{\vk}}$ given by (\ref{probascalar}).  The entropy $S^{(d)}$ is obtained once the fiducial internal energy $U$ is found. It is easily shown to be given by the equivalent form in quantum statistical mechanics
\be U = \sum_{\vk} \frac{\mathcal{E}_k }{e^{\mathcal{E}_{k}}-1}\,.  \label{Ufiduc}\ee Using the identity (\ref{idenN}) and recognizing the following relations
\be \mathcal{E}_k = \ln \Big[\frac{1+N_k}{N_k} \Big]~~;~~ \frac{1}{e^{\mathcal{E}_{k}}-1} = N_k \label{idenqsm}\ee we find the von Neumann entropy
\be S^{(d)} = \sum_{\vk} \Bigg\{(1+N_k)\,\ln(1+N_k) - N_k\,\ln N_k   \Bigg\}\,.  \label{vNS}\ee

\subsection{Interpretation of $S^{(d)}$: entanglement entropy.}\label{subsec:entan}
Consider the full density matrix $\rho_S(\eta)$ eqn. (\ref{rhosout}). Although it describes a pure state, in the out basis this state is a highly correlated, \emph{entangled  state of pairs}, because in this basis the state $\ket{0_I}$ is not a simple product state. Because the members of the particle-anti-particle pairs are correlated, projecting onto a state with $n_{\vk}$ antiparticles of momentum $-\vk$ effectively projects onto the state with $n_{\vk}$ particles with momentum $\vk$. Therefore, consider obtaining a \emph{reduced} density matrix by tracing $\rho_S(\eta)$ over the \emph{anti-particle} states  $\overline{p}$. Because the states $\ket{n_{\vk};\overline{n}_{-\vk}}= \ket{n_{\vk}}\,\ket{\overline{n}_{-\vk}}$ such trace involves terms of the form $(\ket{n_{\vk}}\bra{m_{\vp}})\,\,(\langle \overline{n}_{-\vk}|\overline{m}_{-\vp}\rangle) = (\ket{n_{\vk}}\bra{m_{\vp}}) \,\delta_{\vk,\vp}\,\delta_{n_{\vk},m_{\vp}}$ thereby projecting on particle states diagonal both in number and momentum. Therefore the rapidly varying phases in (\ref{rhosout}) vanish \emph{identically}, yielding

\be \rho^{(r)}_S(\eta) = \mathrm{Tr}_{\overline{p}}\,\rho_S(\eta) =  \Pi_{\vk} \big[1-\tanh^2(\theta_k) \big]  \sum_{n_{\vk}=0}^\infty \Big(\tanh^2(\theta_k)\Big)^{n_{\vk}}\ket{n_{\vk}}\bra{n_{\vk}} \,. \label{reducedrho} \ee Note that because  the density matrix (\ref{diagrho}) is diagonal in the basis of correlated pairs, tracing over one member of the correlated pair, either the particle or the antiparticle keeps the density matrix diagonal with the same probabilities. For example, tracing over the antiparticles reduces (\ref{diagrho}) directly to (\ref{reducedrho}) \emph{with the same eigenvalues, i.e.  probabilities}. This observation is yet another manner to interpret the equivalence with the fiducial quantum statistical mechanical example, now with the fiducial Hamiltonian
\be \widehat{\mathcal{H}}^{(r)} = \sum_{\vk} \mathcal{E}_k\, \widehat{\mathcal{N}}^{(r)}_{\vk} \,,\label{fiduHred}\ee with the \emph{reduced} number operator
\be \widehat{\mathcal{N}}^{(r)}_{\vk} = \sum_{m_{\vk}=0}^\infty m_{\vk}\,\, \ket{m_{\vk}}\bra{m_{\vk}}  \,, \label{pairopred}  \ee namely,
\be \rho^{(r)}_S = \frac{e^{-\widehat{\mathcal{H}}^{(r)}}}{\mathcal{Z}}~~;~~ \mathcal{Z}= \mathrm{Tr}\,e^{-\widehat{\mathcal{H}}^{(r)}} \equiv e^{-\mathbb{F}} \,,\label{rhofidured}\ee with the same $\mathcal{Z}$ and fiducial free energy $\mathbb{F}$ as for $\rho^{(d)}_S$ eqn. (\ref{diagrho}). Hence $\rho^{(r)}_S $ and $\rho^{(d)}_S$ feature the same eigenvalues and yield the same entropy.

The von Neumann entropy associated with  the reduced density matrix $\rho^{(r)}_S(\eta)$, i.e.
\be  S^{(r)} = - \mathrm{Tr}\,\rho^{(r)}_S\,\ln \rho^{(r)}_S \,, \label{redvNS} \ee is the \emph{entanglement entropy}\cite{nielsen}. Therefore, we conclude that   decoherence from rapid dephasing of the off diagonal density matrix elements results in a reduction of the density matrix which is diagonal in the correlated pair basis. This reduction is   identical to tracing over one member of the correlated pair leading to   the entanglement entropy. The equivalence between the entropy resulting from dephasing and decoherence and the entanglement entropy is no accident: it is a direct consequence of the entangled-- correlated-- particle-antiparticle pairs in the out state and that  after decoherence the density matrix is diagonal in this basis of \emph{correlated pairs}. Therefore the diagonal matrix elements, in other words  the probabilities, are exactly the same as when one of the members of the pairs is traced over, which yields the entanglement entropy.
 The result (\ref{vNS})  is remarkably similar to  the quantum kinetic form of the entropy in terms of the distribution function\cite{bernstein}. However, there is an important difference: a complex scalar field has two degrees of freedom, corresponding to particles and antiparticles, therefore if the out state were a superposition   independent single particles and antiparticles we would expect an extra overall factor $2$ multiplying the von Neumann entropy (\ref{vNS}) because of the two independent degrees of freedom. The reason for this discrepancy is that   the density matrix is diagonal in the basis of particle-antiparticle \emph{correlated pairs}, not independent particles and antiparticles. Because of the pairing, for each pair there is effectively   only one degree of freedom, not two as would be the case for independent particles and antiparticles. This is more evident in the identification of the von Neumann entropy with the entanglement entropy which is obtained by tracing over one member of the pairs either particle or antiparticle.

\subsection{Energy density, pressure and entropy.}\label{subsec:total}

During the adiabatic regime and well before matter radiation equality, the decoherence process via dephasing renders the time dependent density matrix in the Schroedinger picture diagonal in the ``out'' basis, namely $\rho^{(d)}_S$. With this density matrix we find
\bea && \mathrm{Tr} \, c^\dagger_{\vk}\, c_{\vk}\,\rho^{(d)}_S = \mathrm{Tr}\, d^\dagger_{\vk} \, d_{\vk}\,\rho^{(d)}_S = \sinh^2(\theta_k) = N_k \,,\nonumber \\ && \mathrm{Tr}\,  c^\dagger_{\vk} \, d^\dagger_{-\vk}\,\rho^{(d)}_S = \mathrm{Tr}\, d_{-\vk} \, c_{\vk}\,\rho^{(d)}_S = 0 \,,  \label{avecd}\eea from which we can now obtain the expectation value of the energy momentum tensor, given by eqn. (\ref{exvalS}) with $\rho_S(\eta) \equiv \rho^{(d)}_S$. The non-vanishing contributions to the expectation values of the expressions (\ref{rhoop},\ref{tmumuope}) are those with terms $c^\dagger c, d^\dagger d$, since the off-diagonal terms of the density matrix $\rho^{(d)}_S$ vanish.

Near matter radiation equality when the dark matter contribution begins to dominate, the adiabatic approximation is very reliable, therefore we keep  the leading order terms in the adiabatic expansions (\ref{capF},\ref{capG}), namely $|\mathcal{F}| = |\mathcal{G}| =1$, yielding
\be \overline{\rho}(\eta) = \mathrm{Tr}\,\widehat{\overline{\rho}}(\eta)\,\rho^{(d)}_S = \frac{1}{2\pi^2\,a^4(\eta)}\,\int_0^{\infty} k^2 \Big[1+ 2 N_k\Big] \omega_k(\eta)\,dk \,, \label{rhodm} \ee
\be \overline{P}(\eta) = \mathrm{Tr}\,\widehat{\overline{P}}(\eta)\,\rho^{(d)}_S = \frac{1}{6\pi^2\,a^4(\eta)}\,\int_0^{\infty} k^2 \Big[1+ 2 N_k\Big] \frac{k^2}{\omega_k(\eta)}\,dk \,.  \label{pressdm} \ee These are precisely the kinetic fluid expressions obtained in ref.\cite{herring} after averaging over the rapid phases in the interference terms. Therefore, this averaging in the energy momentum tensor and the emergence of the kinetic fluid form in the adiabatic regime is a direct manifestation of  decoherence by dephasing in the density matrix, hence also directly related to the emergence of entropy.

The ``1'' inside the brackets in (\ref{rhodm},\ref{pressdm}) correspond to the zero point energy density and pressure. As explained in detail in refs.\cite{herring}, these zero point contributions are subtracted by renormalization of  the energy momentum tensor\cite{bunch,pf,fh,hu,anderson,bir,mottola}. Therefore the contribution from gravitational  particle-antiparticle production to the energy density, pressure and comoving entropy density $ \mathcal{S} = S/V$ ($V$ is comoving volume)  of dark matter are given by the kinetic-fluid forms
\bea  \mathcal{N}_{p\overline{p}} & = &  \frac{1}{\pi^2}\,\int_0^{\infty} k^2   N_k\, dk \label{nbosepp} \\
 \overline{\rho}_{p\overline{p}}(\eta)  & = &  \frac{1}{\pi^2\,a^4(\eta)}\,\int_0^{\infty} k^2   N_k\, \omega_k(\eta)\,dk \label{rhopp} \\
\overline{P}_{p\overline{p}}(\eta) & = &  \frac{1}{3\pi^2\,a^4(\eta)}\,\int_0^{\infty} \frac{k^4}{\omega_k(\eta)}\,   N_k  \,dk \label{Ppp} \\
\mathcal{S}_{p\overline{p}}  & = & \frac{1}{2\pi^2}\, \int_0^{\infty} k^2 \Big[(1+N_k)\,\ln[1+N_k] - N_k\,\ln N_k \Big] \, dk \label{spp}\,, \eea where $\mathcal{N}_{p\overline{p}}$ is the total (particles plus antiparticles ) comoving number density . It is straightforward to confirm covariant conservation
\be \dot{\overline{\rho}}_{p\overline{p}}(t)+ 3\,\frac{\dot{a}}{a}\,\Big(\overline{\rho}_{p\overline{p}}(t)+\overline{P}_{p\overline{p}}(t) \Big) =0 \,, \label{covacons}\ee along with the conservation of the comoving entropy density
\be \dot{\mathcal{S}}_{p\overline{p}}=0 \,,\label{scons}\ee
where the dot stands for derivative with respect to comoving time. Although the comoving entropy density is proportional (up to a factor 2) to the quantum kinetic expression, it is not to be identified with a thermodynamic entropy, as shown above it is the entanglement entropy resulting from the loss of information as a consequence of dephasing and decoherence from the interference between particle and antiparticle out states. The equivalence with the entanglement entropy is a consequence of the correlations in the particle-antiparticle pairs, tracing over one member is equivalent to neglecting the off-diagonal matrix elements.

 The result (\ref{spp}) is similar to the expression for the entanglement entropy obtained    in ref.\cite{beilok} for bosonic particle production after tracing one member of the produced pairs from the Wigner distribution function. While in this reference the tracing over one member of the pairs was carried out to obtain the entanglement entropy, we emphasize that in our case, the main origin of entropy is the decoherence via dephasing during the adiabatic regime. The fact that this entropy is exactly the same as the entanglement entropy is an \emph{a posteriori} conclusion on the equivalence between the entropy emerging from the decoherence via dephasing and the entanglement entropy.

\subsection{Entropy for ultra light dark matter:}\label{subsec:uldment}
In ref.\cite{herring} the case of gravitationally produced ultra light dark matter has been studied under the same conditions assumed in this article. In this reference it was established that a scalar field minimally coupled to gravity and with mass $m \simeq 10^{-5}\,\mathrm{eV}$ yields the correct dark matter abundance and is a cold dark matter candidate with a very small free streaming length. The distribution function is given by equation (\ref{nofkmcfin}). It features an infrared enhancement $\propto 1/k^3$ and the large factor $H_{dS}/m \gg 1$, both consequences of a light scalar minimally coupled to gravity during inflation. Since $D(z) \simeq 1/z$ for $z\gg 1$ the occupation number $N_k \gg 1$ in  the region $0 \leq z \ll \sqrt{H_{dS}/m}$.

 The comoving number density of gravitationally produced cold dark matter scalar particles   has been obtained in ref.\cite{herring}, it is given by
\be \mathcal{N}_{p\overline{p}} \simeq  \bigg( \frac{H_{dS}}{4\,\pi\,m}\bigg)^2\, \, \Big(2mH_R \Big)^{3/2}\, D(0)\,\ln\Big[ \frac{\sqrt{2mH_R}}{H_0}\Big]\,.  \label{densbos}\ee

The leading contribution to the comoving entropy density (\ref{spp}) can be extracted by implementing the following steps: a)  changing integration variable to $z$ given by (\ref{zdef}) b)  taking the limit $N_k \gg 1$ in the region of integration dominated by the infrared $0 \leq z \leq z_c$ where $1\ll z_c \ll \sqrt{H_{dS}/m}$,  yielding
\be \mathcal{S}_{p\overline{p}}   \simeq     \frac{\big(2\,m\,H_R \big)^{3/2}}{2\pi^2}\, \int^{z_c}_0   z^2 \Big[ \ln(N_k) +\cdots \Big]  \, dz  \,,\label{spp2} \ee where the dots stand for subleading terms  of order $1/N_k$ for $N_k \gg 1$. It is more instructive to obtain the dimensionless \emph{specific entropy}, namely the entropy per particle $\mathcal{S}_{p\overline{p}}/\mathcal{N}_{p\overline{p}}$. To leading order in $H_{dS}/m\gg 1 $ we find
\be \frac{\mathcal{S}_{p\overline{p}}}{\mathcal{N}_{p\overline{p}}} \simeq \frac{16}{3\,D(0)}\, \,\frac{\ln\big(H_{dS}/m \big)\,z^3_c}{\Big(\frac{H_{dS}}{m}\Big)^2\,\ln\Big[ \frac{\sqrt{2mH_R}}{H_0}\Big]} \Bigg\{1- \frac{1}{2\,\ln\big(H_{dS}/m \big)} \,\Big[\ln(8\sqrt{2})-(4/3-4\ln z_{c})-\frac{0.17}{z_{c}^{3}} \Big]  \Bigg\}   \,.\label{specS} \ee  For ultra light dark matter with $H_0 \ll m \ll H_{dS} $  (for example with $H_{dS} = 10^{13}\,\mathrm{GeV}, m \simeq 10^{-5}\,\mathrm{eV}$)  it follows that the specific entropy
\be \frac{\mathcal{S}_{p\overline{p}}}{\mathcal{N}_{p\overline{p}}} \ll 1 \,. \label{smalss}\ee

A large occupation number in an narrow momentum region and with a very small specific entropy are all hallmarks of a \emph{condensed state}, these are precisely the conditions of a Bose Einstein Condensate. However, in this case of gravitationally produced particles, this is not a condensate in the usual manner because the expectation value of the field vanishes, therefore it is not described by a coherent state. Instead this a condensed state of correlated pairs entangled in momentum but of total zero momentum in a two-mode squeezed state\cite{barnett}.

For a value of the mass that yields the correct dark matter abundance, $m \simeq 10^{-5}\,\mathrm{eV}$\cite{herring}, the ratio of the comoving dark matter entropy $\mathcal{S}_{p\overline{p}}$ to that of the (CMB)
\be \mathcal{S}_{cmb} \simeq T^3_0~~;~~T_0 \simeq 10^{-4} \,\mathrm{eV}\,\label{encmb}\ee    yields,
\be \frac{\mathcal{S}_{p\overline{p}}}{\mathcal{S}_{cmb}} \simeq 10^{-45}\,, \label{ratiocmbbos}\ee therefore, if ultra light dark matter is gravitationally produced,  the entropy of the Universe today is dominated by the (CMB).

\section{Fermionic Dark Matter}\label{sec:fdm}

The results obtained above for a complex scalar are, in fact, much more general and apply with few modifications primarily due to the different statistics,  to the case of gravitationally produced fermionic dark matter. We analyze this case by briefly summarizing the results of ref.\cite{herringfer} to which we refer the reader for a more comprehensive treatment.

In comoving
coordinates, the action for a Dirac field is given by
\be
S   =  \int d^3x \; dt \;  \sqrt{-g} \,  \overline{\Psi}  \Big[i\,\gamma^\mu \;  \mathcal{D}_\mu -m  \Big]\Psi     \,. \label{lagrads}
\ee

Introducing  the vierbein field $e^\mu_a(x)$  defined as
$$
g^{\mu\,\nu}(x) =e^\mu_a (x)\;  e^\nu_b(x) \;  \eta^{a b} \; ,
$$
\noindent where $\eta_{a b}= \mathrm{diag}(1,-1,-1,-1)$ is the Minkowski space-time metric,
the curved space time Dirac gamma- matrices $\gamma^\mu(x)$  are given
by
\be
\gamma^\mu(x) = \gamma^a e^\mu_a(x) \quad , \quad
\{\gamma^\mu(x),\gamma^\nu(x)\}=2 \; g^{\mu \nu}(x)  \; ,
\label{gamamtx}\ee where the $\gamma^a$ are the Minkowski space time Dirac matrices.

 The fermion covariant derivative $\mathcal{D}_\mu$ is given in terms of the spin connection by\cite{weinbergbook,casta,parkerbook,birrell}

\be
\mathcal{D}_\mu   =    \partial_\mu + \frac{1}{8} \;
[\gamma^c,\gamma^d] \;  e^\nu_c  \; \left(\partial_\mu e_{d \nu} -\Gamma^\lambda_{\mu
\nu} \;  e_{d \lambda} \right) \,,  \label{fermicovader}
\ee  where $\Gamma^\lambda_{\mu
\nu}$ are the usual Christoffel symbols.

For a spatially flat Friedmann-Robertson-Walker cosmology
  in conformal time  with  metric is given by eqn. (\ref{metric})
  the vierbeins  can be obtained easily.  Introducing the conformally rescaled fields
\be  a^{\frac{3}{2}}(\eta)\,{\Psi(\vx,t)}= \psi(\vx,\eta)\,, \label{rescaledfields}\ee
  the action becomes
   \be  S    =
  \int d^3x \; d\eta \,   \overline{\psi} \;  \Big[i \;
{\not\!{\partial}}- M(\eta)   \Big]
 {\psi}    \;, \label{rescalagds}\ee
  with
   \be M (\eta) = m  \,a(\eta) \,,  \label{masfer}\ee and the $\gamma^a$ matrices are the usual Minkowski space time ones taken to be in the standard Dirac representation. We consider the fermion mass $m$ much smaller than the Hubble scale  during inflation, namely $m/H_{dS}\ll 1$ but otherwise arbitrary.

 The Dirac equation for the conformally rescaled fermi field becomes
\be  \Big[i \;
{\not\!{\partial}}- M(\eta)    \Big]
 {\psi}  = 0\,,    \label{diraceqn}\ee   and expand $ \psi({\vec x},\eta) $  in a comoving volume $V$  as
\be
\psi(\vec{x},\eta) =    \frac{1}{\sqrt{V}}
\sum_{\vec{k},s}\,   \left[b_{\vec{k},s}\, U_{s}(\vec{k},\eta) +
d^{\dagger}_{-\vec{k},s}\, V_{s}(-\vec{k},\eta)
 \right]\,e^{i \vec{k}\cdot\vec{x}} \; ,
\label{psiex}
\ee
  and  the spinor mode functions $U,V$ obey the  Dirac equations
\bea
&& \Bigg[i \; \gamma^0 \;  \partial_\eta - \vec{\gamma}\cdot \vec{k}
-M(\eta) \Bigg]U_s(\vec{k},\eta)   =  0 \label{Uspinor} \\
&& \Bigg[i \; \gamma^0 \;  \partial_\eta - \vec{\gamma}\cdot \vec{k} -M(\eta)
\Bigg]V_s(-\vec{k},\eta)   =   0 \,. \label{Vspinor}
\eea

Finally, the spinor solutions  are given by\cite{herringfer}

\be U_s(\vec{k},\eta) = N\,\left( \begin{array}{c}
                                     \mathcal{F}_k(\eta)\, \xi_s\\
                                    k\,f_k(\eta) \, s \, \xi_s
                                  \end{array}\right)\,,  \label{Uspin}\ee

 \be V_s(-\vec{k},\eta) = N\,\left( \begin{array}{c}
                                      -k\,f^*_k(\eta) \, s\,\xi_s\\
                                    \mathcal{F}^*_k(\eta)  \,   \xi_s
                                  \end{array}\right)\,,  \label{Vspin}\ee  where

 \be \mathcal{F}_k(\eta) =    if'_k(\eta)+M(\eta) f_k(\eta)\,, \label{capFfer}\ee  and the functions $f_k(\eta)$ are solutions of\cite{herringfer}
 \be \left[\frac{d^2}{d\eta^2} +
k^2+M^2 (\eta)-i \; M' (\eta)\right]f_k(\eta)   =  0 \,, \label{feq}\ee with ``in'' boundary conditions \be f_k(\eta) ~\rightarrow ~e^{-ik\eta}\,, \label{inbc} \ee as $\eta \rightarrow -\infty$ during inflation\cite{herringfer}. The two component spinors $\xi_s$ are helicity eigenstates, namely
 \be \vec{\sigma}\cdot\vec{k} = s \, k \, \xi_s ~~;~~ s = \pm 1\,,  \label{helicity} \ee
 and $N$ is a (constant) normalization factor.

                                   The spinor solutions are normalized as follows
 \be    U^\dagger_s(\vec{k},\eta)\,  U_{s'}(\vec{k},\eta) =\delta_{s,s'}~~;~~  V^\dagger_s(-\vec{k},\eta)\,  V_{s'}(-\vec{k},\eta) =\delta_{s,s'}\,,\label{normas}\ee yielding
 \be |N|^2\Big[\mathcal{F}^*_k(\eta)\,\mathcal{F}_k(\eta)+ k^2 f^*_k(\eta)\, f_k(\eta) \Big] = 1\,. \label{normaN} \ee With these normalization conditions the operators $b_{\vec{k},s},d_{\vec{k},s}$ in the field expansion (\ref{psiex})  obey the usual canonical anticommutation relations.

 Furthermore, it is straightforward to confirm that
\be U^\dagger_s(\vec{k},\eta)\,  V_{s'}(-\vec{k},\eta) =0 ~~~\forall s,s' \,. \label{ortho}\ee

The spinors $U_s,V_s$ furnish a complete set of four   independent solutions of the Dirac equation.

During the inflationary stage, considered as an  spatially flat de Sitter space-time, the functions $f_k$ obey
\be \Bigg[ \frac{d^2}{d\tau^2} + k^2 - \frac{\nu^2-1/4}{\tau^2}\Bigg]f_k(\tau) = 0 ~~;~~ \tau = \eta-2\eta_R~~;~~ \nu = \frac{1}{2}+i\,\frac{m}{H_{dS}}\,. \label{desieqn} \ee   The solution with ``in'' boundary conditions (\ref{inbc}) is given by
\be f_k(\tau) = \sqrt{-\frac{\pi k \tau}{2}}~e^{i\pi(\nu+1/2)/2}~H^{(1)}_\nu(-k\tau) \,, \label{fdS}\ee where $H^{(1)}_{\nu}$ is a Hankel function. The operators $b_{\vk,s},d_{\vk,s}$ in the field expansion (\ref{psiex}) are chosen to annihilate the ``in'' vacuum state $\ket{0_I}$, namely
\be b_{\vk,s}\ket{0_I} =0 ~~;~~ d_{\vk,s}\ket{0_I} =0\,, \label{invac}\ee with the mode functions $f_k$ given by (\ref{fdS}), the state $\ket{0_I}$ corresponds to the Bunch-Davies vacuum.

 Since we are considering an instantaneous transition between inflation and radiation domination, and because the Dirac equation is first order in time, the matching conditions correspond to the continuity of the spinor wave functions across the transition.

Defining $\psi^<(\vec{x},\eta)$ and $\psi^>(\vec{x},\eta)$ the fermion field for $\eta < \eta_R$ (inflation) and
$\eta > \eta_R$ (RD) respectively,  the matching condition   is
\be \psi^<(\vec{x},\eta_R) = \psi^>(\vec{x},\eta_R)\,.\label{match} \ee
This continuity condition along with the continuity of the scale factor and Hubble rate at $\eta_R$  results in that the energy density is \emph{continuous at the transition}\cite{herringfer}.

Introducing the Dirac spinors during the inflationary ($\eta < \eta_R$)  and (RD) ($\eta > \eta_R$) stages as $U^<\,,\,V^<$  and   $U^>\,,\,V^>$ respectively,  it follows from the matching condition (\ref{match}) that
\bea &&  U^<_{s}(\vec{k},\eta_R) = U^>_{s}(\vec{k},\eta_R)\,, \label{Umach}\\
&&  V^<_{s}(-\vec{k},\eta_R) = V^>_{s}(-\vec{k},\eta_R)\,. \label{Vmach}\eea

We define the mode functions during (RD) as $h_k(\eta)$ to distinguish them from the solutions (\ref{fdS}) during inflation. These obey the mode equations
\be \Bigg[ \frac{d^2}{d\eta^2}+\omega^2_k(\eta)   - i\,m H_R  \Bigg]h_k(\eta) = 0 ~~;~~\omega^2_k(\eta)= k^2 + m^2 H^2_R \eta^2 \,.\label{heqn} \ee Similarly to the spinor solutions (\ref{Uspin},\ref{Vspin}) we now find

\be  \mathcal{U}_s(\vec{k},\eta) = \widetilde{N}\,\left( \begin{array}{c}
                                     \mathcal{H}_k(\eta)\, \xi_s\\
                                    k\,h_k(\eta) \, s \, \xi_s
                                  \end{array}\right)\,,  \label{Uspinrd}\ee

 \be \mathcal{V}_s(-\vec{k},\eta) = \widetilde{N}\,\left( \begin{array}{c}
                                      -k\,h^*_k(\eta) \, s\,\xi_s\\
                                    \mathcal{H}^*_k(\eta)  \,   \xi_s
                                  \end{array}\right)\,,  \label{Vspinrd}\ee  where we have introduced
\be \mathcal{H}_k(\eta) =  ih'_k(\eta)+M(\eta) h_k(\eta)\,,   \label{calH}\ee   and
                                  $\widetilde{N}$ is a (constant) normalization factor chosen  so that
 \be     \mathcal{U}^\dagger_s(\vec{k},\eta)\,  \mathcal{U}_{s'}(\vec{k},\eta) =\delta_{s,s'}~~;~~  \mathcal{V}^\dagger_s(-\vec{k},\eta)\,  \mathcal{V}_{s'}(-\vec{k},\eta) =\delta_{s,s'}\,,\label{normasrd}\ee  yielding
 \be | \widetilde{N}|^2 \Big[ \mathcal{H}^*_k(\eta) \mathcal{H}_k(\eta)+ k^2 h^*_k(\eta)h_k(\eta) \Big] =1\,.  \label{nortilN}\ee
Again,  it is straightforward to confirm that
\be \mathcal{U}^\dagger_s(\vec{k},\eta)\,  \mathcal{V}_{s'}(-\vec{k},\eta) =0 \,. \label{orthord}\ee

The mode equation (\ref{heqn}) admits a solution of the form\cite{herringfer} (see appendix (\ref{app:ferad}))
\be h_k(\eta) = e^{-i\int^\eta \Omega_k(\eta')d\eta'}\,, \label{soluh1}\ee where $\Omega_k(\eta)$ obeys a differential equation that can be systematically solved in the adiabatic expansion and is analyzed in appendix (\ref{app:ferad}). It relies on the ratio $H(\eta)/m \ll 1$ which during the (RD) era implies that $a(\eta) \gg 10^{-17}/\sqrt{m(\mathrm{eV})}$, for the value $m\simeq 10^8\,\mathrm{GeV}$ which saturates the dark matter bound as found in ref.\cite{herringfer},  its range of validity begins well before matter radiation equality at $a_{eq} \simeq 10^{-4}$. We choose the solution of (\ref{heqn}) to feature the asymptotic ``out'' boundary condition
\be h_k(\eta) \rightarrow  e^{-i\int^\eta \omega_k(\eta') \,d\eta'} \,. \label{hout}\ee

 With this boundary condition, the spinor solutions during the (RD) era (\ref{Uspinrd},\ref{Vspinrd}) satisfy  the  asymptotic ``out''  boundary conditions
\be  \mathcal{U}_s(\vec{k},\eta) \rightarrow \propto \,  e^{-i\int^\eta \omega_k(\eta') \,d\eta'}~~;~~ \mathcal{V}_s(\vec{k},\eta) \rightarrow \propto \, e^{i\int^\eta \omega_k(\eta') \,d\eta'}\,.  \label{rdbcs}\ee    therefore describing ``out'' particle and anti-particle solutions with helicities $\pm 1$,  defining a  complete set of four solutions of the Dirac equation  during (RD).

It is convenient to introduce the following dimensionless combinations,
\be z= \sqrt{m H_R}\,\, \eta ~~;~~ q = \frac{k}{\sqrt{m H_R}} ~~;~~ \lambda = q^2 - i \label{dimcombos}\ee in terms of which eqn. (\ref{heqn}) becomes
\be \frac{d^2}{dz^2} h_k(z) + (z^2 + \lambda)h_k(z) =0 \,, \label{dimeqn}\ee the solutions of which are the parabolic cylinder functions\cite{gr,as,nist,bateman,magnus}
\be D_\alpha[\sqrt{2} e^{i\pi/4} z]~~;~~ D_\alpha[\sqrt{2} e^{3i\pi/4} z]~~;~~\alpha = -\frac{1}{2}-i \, \frac{\lambda}{2} = -1 - i\,\frac{q^2}{2}\,.  \label{solus}\ee The solution that fulfills the ``out'' boundary condition (\ref{hout}) (see appendix A in ref.\cite{herringfer}) is given by
\be h_k(\eta) =  D_\alpha[\sqrt{2} e^{i\pi/4} z] \,. \label{soluh}\ee The general solution for the spinor wave functions $U^>,V^>$ during the (RD) era are linear combinations of the four independent solutions (\ref{Uspinrd},\ref{Vspinrd}). In principle, with four independent solutions during inflation matching onto four independent solutions during (RD) there would be a $4\times 4$ matrix of Bogoliubov coefficients,  however,  because helicity is conserved, the linear combinations are given by
\bea &&  U^>_s(\vec{k},\eta) = A_{k,s}\, \mathcal{U}_s(\vec{k},\eta)+ B_{k,s} \,\mathcal{V}_s(-\vec{k},\eta)\label{Ugreat} \\
&& V^>_s(-\vec{k},\eta) = C_{k,s}\, \,\mathcal{V}_s(-\vec{k},\eta)+
D_{k,s}\, \mathcal{U}_s(\vec{k},\eta) \,.  \label{Vgreat} \eea The  Bogoliubov coefficients $A_{k,s} \cdots D_{k,s}$ are obtained from the matching conditions (\ref{Umach},\ref{Vmach}) and the relations (\ref{normasrd},\ref{orthord}). These   obey the   relations\cite{herringfer}
\be D_{k,s} = - B^*_{k,s} ~~;~~ C_{k,s} = A^*_{k,s} \,, \label{bogorels}\ee and
\be |A_{k,s}|^2 + |B_{k,s}|^2 = 1\,.  \label{bogferun}\ee

During the (RD) era, with $U_s \equiv U^>_s; V_s \equiv V^>_s$ with $U^>, V^>$ given by (\ref{Ugreat},\ref{Vgreat}) the field expansion (\ref{psiex}) in terms of the spinor solutions with out boundary conditions (\ref{rdbcs}) becomes

 \be
\psi(\vec{x},\eta) =    \frac{1}{\sqrt{V}}
\sum_{\vec{k},s}\,   \left[\widetilde{b}_{\vec{k},s}\, \mathcal{U}_{s}(\vec{k},\eta) +
\widetilde{d}^{\,\dagger}_{-\vec{k},s}\, \mathcal{V}_{s}(-\vec{k},\eta)
 \right]\,e^{i \vec{k}\cdot\vec{x}} \; ,
\label{psiexrd}
\ee where
\bea \widetilde{b}_{\vec{k},s} & = &   {b}_{\vec{k},s} A_k + {d}^{\dagger}_{-\vec{k},s} D_{k,s} \label{btil} \\
\widetilde{d}^{\,\dagger}_{-\vec{k},s} & = &   {d}^{\dagger}_{-\vec{k},s} C_{k,s} + {b}_{\vec{k},s} B_{k,s} \,. \label{ddtil}
\eea The relations (\ref{bogorels},\ref{bogferun}) imply that the new operators $\widetilde{b},\widetilde{d}$ obey   canonical anticommutation relations. The operators $\widetilde{b}^\dagger$ and $\widetilde{d}^\dagger$ create asymptotic particle and antiparticle states respectively.  In particular we find that the number of asymptotic ``out'' particle and antiparticle states in the Bunch-Davies vacuum state (\ref{invac}) are the same and given by
\be  \langle 0_I| \widetilde{b}^\dagger_{\vec{k},s}   \widetilde{b}_{\vec{k},s}|0_I\rangle    =   |D_{k,s}|^2 = \langle 0_I| \widetilde{d}^\dagger_{-\vec{k},s}   \widetilde{d}_{-\vec{k},s}|0_I\rangle    =    |B_{k,s}|^2 \,. \label{partsfer}\ee We identify the number of ``out'' particles,  equal the number of ``out'' anti-particles as
\be \langle 0_I| \widetilde{b}^\dagger_{\vec{k},s}   \widetilde{b}_{\vec{k},s}|0_I\rangle  =  \langle 0_I| \widetilde{d}^\dagger_{-\vec{k},s}   \widetilde{d}_{-\vec{k},s}|0_I\rangle =
|B_{k,s}|^2 \equiv N_k \label{nkfer}\ee  with $N_k= |B_{k,s}|^2$ being the \emph{distribution function of produced particles and antiparticles}. The relation (\ref{bogferun}) implies that
\be |B_{k,s}|^2 \leq 1\,, \label{maxn} \ee for each helicity $s$,  consistent with Pauli exclusion. For $m \ll H_{dS}$ it is found in ref.(\cite{herringfer})  that
\be   N_k=  |B_{k,s}|^2 = \frac{1}{2} \Big[1-\big(1-e^{-\frac{k^2}{2mT_H}}\big)^{1/2}\Big]\,,  \label{almMB}\ee in terms of the emergent temperature\cite{herringfer}
\be   T_H = \frac{H_R}{2\pi} \simeq 10^{-36}\,\mathrm{eV}\, . \label{Thawk}\ee

In the adiabatic regime during (RD) the spinors  $ \mathcal{U}_s(\vec{k},\eta),\mathcal{V}_s(-\vec{k},\eta)$ can be written as (see appendix (\ref{app:ferad}) and ref.\cite{herringfer})

\be \mathcal{U}_s(\vec{k},\eta) =  {e^{-i\int^\eta_{\eta_i} \omega_k(\eta')\,d\eta'}}\,\widetilde{\mathcal{U}}_s(\vec{k},\eta)~~;~~ \mathcal{V}_s(-\vec{k},\eta) =  {e^{i\int^\eta_{\eta_i} \omega_k(\eta')\,d\eta'}} \,\widetilde{\mathcal{V}}_s(-\vec{k},\eta)\,, \label{fastslowfer}\ee where $\widetilde{\mathcal{U}}_s(\vec{k},\eta)~;~\widetilde{\mathcal{V}}_s(-\vec{k},\eta)$ are slowly varying functions of time during this regime, and again $\eta_i$ is some early time in the adiabatic regime. To leading (zeroth) order in the adiabatic expansion these are given by (see appendix (\ref{app:ferad}))
\be \widetilde{\mathcal{U}}_s(\vec{k},\eta) = \frac{1}{\Big[2\omega_k(\eta)(\omega_k(\eta)+M(\eta))\Big]^{1/2}} ~~\left( \begin{array}{c}
                                     (\omega_k(\eta)+M(\eta) )\, \xi_s\\
                                    k \, s \, \xi_s
                                  \end{array}\right) \,,  \label{Uspinrdasy}\ee

 \be \widetilde{\mathcal{V}}_s(-\vec{k},\eta) =  \frac{1}{\Big[2\omega_k(\eta)(\omega_k(\eta)+M(\eta))\Big]^{1/2}}~~ \left( \begin{array}{c}
                                      -k \, s\,\xi_s\\
                                    (\omega_k(\eta) + M(\eta))  \,   \xi_s
                                  \end{array}\right) \,.  \label{Vspinrdasy}\ee

 \subsection{Energy density,  pressure and entropy:}\label{subsec:EP}

 The energy momemtum tensor for Dirac fields is given by \cite{parkerbook,rio,barbero,landete}
 \be T^{\mu \nu} = \frac{i}{2} \Big( \overline{\Psi} \gamma^\mu  \stackrel{\leftrightarrow}{\mathcal{D}^\nu}   \,\Psi \Big) + \mu \leftrightarrow \nu \label{tmunudirac} \ee

 In terms of conformal time and the conformally rescaled fields (\ref{rescaledfields})
 the energy density $\rho$  and pressure  $P$ as operators are given by
 \be \widehat{\rho}(\vx,\eta) =   T^0_0(\vx,\eta)  = \frac{i}{ 2 a^4(\eta)} ~     \Big(\psi^\dagger (\vec{x},\eta) \frac{d}{d\eta}\,\psi (\vec{x},\eta) - \frac{d}{d\eta}\,\psi^\dagger (\vec{x},\eta) \,\psi(\vec{x},\eta) \Big)   \,, \label{rhoopfer}\ee

 \be \widehat{P}(\vx,\eta) =  -\frac{1}{3} \sum_{j}  T^j_j (\vx,\eta)   = \frac{-i}{6 a^4(\eta)} ~   \Big(\psi^\dagger (\vec{x},\eta) \, \vec{\alpha}\cdot \vec{\nabla} \,\psi (\vec{x},\eta)-\vec{\nabla}\psi^\dagger (\vec{x},\eta)\cdot \vec{\alpha} \,\psi (\vec{x},\eta)\Big)   \,, \label{Popfer}\ee

  The expectation value of the energy momentum tensor in the Bunch-Davies vacuum state is given by
 \be \langle 0_I|T^\mu_\nu|0_I\rangle = \mathrm{diag}\big(\rho(\eta),-P(\eta),-P(\eta),-P(\eta)\big) \,, \label{EMTdiag}\ee only the homogeneous and isotropic component of the energy momentum tensor contributes to the expectation value. Because we want to extract the rapid time dependence during the adiabatic era, we obtain this homogeneous component by averaging the above operators in the comoving volume $V$,
 just as in the bosonic case we obtain

 \be \frac{1}{V} \int  d^3 x \, T^{0}_{0}(\vx,\eta)  = \widehat{\overline{\rho}}(\eta)~~;~~ -\frac{1}{3\,V} \int  d^3 x  \, \sum_{j}  T^j_j (\vx,\eta) =  \widehat{\overline{P}}(\eta)\,.\label{volavefer}\ee During the (RD) era and near matter radiation equality when the adiabatic approximation becomes very reliable, we obtain these operators by expanding the fermionic field in  the ``out'' basis as in eqn. (\ref{psiexrd}), and writing the spinors as in eqn. (\ref{Uspinrdasy},\ref{Vspinrdasy}) separating the fast phases from the slowly varying spinors $\widetilde{U},\widetilde{V}$. We find
 \bea   \widehat{\rho}(\eta)  & = &  \overline{\rho}_{vac}(\eta)+ \widehat{\overline{\rho}}_{int}(\eta)+ \widehat{\overline{\rho}}_{pp}(\eta)\label{3rhos}\\
  \widehat{P}(\eta)  & = &  \overline{P}_{vac}(\eta)+ \widehat{\overline{P}}_{int}(\eta)+ \widehat{\overline{P}}_{pp}(\eta)\label{3Ps}\,, \eea with
  \bea && \overline{\rho}_{vac}    =      \frac{1}{ V\, a^4(\eta)} ~    \sum_{\vk;s=\pm 1}\, \Big[\widetilde{\mathcal{V}}^\dagger_s(-\vec{k},\eta)\,\Sigma(\vec{k},\eta)\, \,\widetilde{\mathcal{V}}_s(-\vec{k},\eta)  \Big]\,  \,,\label{rhovac} \\
&& \widehat{\overline{\rho}}_{int}    =       \frac{1}{V\, a^4(\eta)} ~    \sum_{\vk;s=\pm 1}\,  \Big[ \widetilde{d}_{-\vk,s}\,\widetilde{b}_{\vk,s}\,e^{-2i\int^\eta_{\eta_i} \omega_k(\eta')\,d\eta'}\, \,\widetilde{\mathcal{V}}^\dagger_s(-\vec{k},\eta)\,\Sigma(\vec{k},\eta)\, \,\widetilde{\mathcal{U}}_s(\vec{k},\eta) +h.c. \Big]\,  \,,\label{rhoint}\\
&&  \widehat{\overline{\rho}}_{p\overline{p}}   =       \frac{1}{V\,  a^4(\eta)} ~    \sum_{\vk;s=\pm 1}\,  \Big[\widetilde{b}^\dagger_{\vk,s}\widetilde{b}_{\vk,s}\widetilde{\mathcal{U}}^\dagger_s(\vec{k},\eta)\,
\Sigma(\vec{k},\eta)\, \,\widetilde{\mathcal{U}}_s(\vec{k},\eta)
      -\widetilde{d}^\dagger_{-\vk,s}\widetilde{d}_{-\vk,s}\widetilde{\mathcal{V}}^\dagger_s(-\vec{k},\eta)\,\Sigma(\vec{k},\eta)\, \,\widetilde{\mathcal{V}}_s(-\vec{k},\eta)  \Big]\,  \,,\nonumber \\\label{rhopp1}   \eea where
 \be \Sigma(\vec{k},\eta)=\vec{\alpha}\cdot \vec{k} + \gamma^0 M(\eta)\,,  \label{enesig}\ee is the conformal time instantaneous Dirac Hamiltonian, and

  \bea \overline{P}_{vac} & = &   \frac{1}{3 \,V\, a^4(\eta)} ~    \sum_{\vk;s=\pm 1}\, \Big[\widetilde{\mathcal{V}}^\dagger_s(-\vec{k},\eta)\,\big(\vec{\alpha}\cdot \vec{k} \big)\, \,\widetilde{\mathcal{V}}_s(-\vec{k},\eta)  \Big]\,  \,,\label{Pvac} \\
\widehat{\overline{P}}_{int} & = &    \frac{1}{3\,V\, a^4(\eta)} ~   \sum_{\vk;s=\pm 1}\,  \Big[\widetilde{d}_{-\vk,s}\,\widetilde{b}_{\vk,s}\,e^{-2i\int^\eta_{\eta_i} \omega_k(\eta')\,d\eta'}\, \,\widetilde{\mathcal{V}}^\dagger_s(-\vec{k},\eta)\,\big(\vec{\alpha}\cdot \vec{k} \big) \,\widetilde{\mathcal{U}}_s(\vec{k},\eta) +h.c. \Big]\,  \,,\label{Pint}\\
 \widehat{\overline{P}}_{p\overline{p}} & = &   \frac{1}{3 \, V\,  a^4(\eta)} ~    \sum_{\vk;s=\pm 1}\,  \Big[\widetilde{b}^\dagger_{\vk,s}\widetilde{b}_{\vk,s}\widetilde{\mathcal{U}}^\dagger_s(\vec{k},\eta)
 \,\big(\vec{\alpha}\cdot \vec{k} \big) \,\widetilde{\mathcal{U}}_s(\vec{k},\eta)
 -  \widetilde{d}^\dagger_{-\vk,s}\widetilde{d}_{-\vk,s}\widetilde{\mathcal{V}}^\dagger_s(-\vec{k},\eta)
 \,\big(\vec{\alpha}\cdot \vec{k} \big) \,\widetilde{\mathcal{V}}_s(-\vec{k},\eta)  \Big]\,    \,. \nonumber \\
 \label{Ppp1}    \eea
 $\overline{\rho}_{vac};\overline{P}_{vac}$ are the zero point (``out'' vacuum) contributions to the energy density and pressure. The terms $\widehat{\overline{\rho}}_{int};\widehat{\overline{P}}_{int}$ feature the fast oscillations associated with the interference between particle and antiparticles similar to the complex bosonic case studied above. As discussed in the previous section, these oscillations average out on comoving time scales equal to or shorter than $\simeq 1/m \ll 1/H(t)$ leaving only the slowly varying contributions $\overline{\rho}_{vac},\overline{\rho}_{p\overline{p}}~;~\overline{P}_{vac},\overline{P}_{p\overline{p}}$. Following the same strategy as in the bosonic case, we introduce the zeroth-order adiabatic Hamiltonian,
 \be H_0(\eta) = \sum_{\vk;s} \Big[ \widetilde{b}^\dagger_{\vk,s}\widetilde{b}_{\vk,s}+\widetilde{d}^\dagger_{\vk,s}\widetilde{d}_{\vk,s}
 \Big] \,\omega_k(\eta)~~;~~ \Big[H_0(\eta),H_0(\eta') \Big]=0\,\,\,\forall \eta,\eta'\,, \label{H0fer}\ee and the time evolution operator
 \be U_0(\eta,\eta_i) = e^{-i\int^\eta_{\eta_i} H_0(\eta')\,d\eta'} \,, \label{U0fer}\ee from which it follows that
 \be U^{-1}_0(\eta,\eta_i)~ \widetilde{b}_{\vk,s} ~ U_0(\eta,\eta_i) = \widetilde{b}_{\vk,s}\,e^{-i\int^{\eta}_{\eta_i} \omega_k(\eta') d\eta'} ~~;~~ U^{-1}_0(\eta,\eta_i)~  \widetilde{d}_{\vk,s} ~ U_0(\eta,\eta_i) = \widetilde{d}_{\vk,s}\,e^{-i\int^{\eta}_{\eta_i} \omega_k(\eta') d\eta'} \,.\label{bdofetafer}\ee It is clear that the fermionic case is very similar to that of the complex scalar case studied in the previous section with the important difference in the statistics. Following the steps described for the scalar  case, we define the  Schroedinger picture fermion operator during the adiabatic regime in the (RD) era
 \be \psi(\vx,\eta) = U_0(\eta,\eta_i)\,\psi_S(\vx,\eta) \, U^{-1}_0(\eta,\eta_i) \,,\ee with
 \be \psi_S(\vx,\eta) =  \frac{1}{\sqrt{V}}
\sum_{\vec{k},s}\,   \left[\widetilde{b}_{\vec{k},s}\, \widetilde{\mathcal{U}}_{s}(\vec{k},\eta) +
\widetilde{d}^{\,\dagger}_{-\vec{k},s}\, \widetilde{\mathcal{V}}_{s}(-\vec{k},\eta)
 \right]\,e^{i \vec{k}\cdot\vec{x}} \; ,
\label{psiexrdfer}   \ee this field evolves slowly in time in the adiabatic regime. A similar definition of Schroedinger picture operators is carried out for the energy momentum tensor just as in the complex scalar case. The density matrix   evolved in time in the Schroedinger picture is given by equation (\ref{densmat}). In appendix (\ref{app:bogofer})) we show  that the fermionic ``in'' Bunch-Davies vacuum state $\ket{0_I}$ is now given in terms of the out states by

\be \ket{0_I} = \Pi_{\vk,s}\Bigg\{\Big[\cos(\theta_k)\Big]~ \,\sum_{n_{\vk,s}=0}^{1} \Big(-e^{2i\varphi_-(k)}\,\tan(\theta_k) \Big)^{n_{\vk,s}} \ket{n_{\vk,s};\overline{n}_{-\vk,s}}\Bigg\} \,, \label{vacin1}\ee
  the fermionic ``out'' particle-antiparticle pair states are given by
 \be  \ket{n_{\vk,s};\overline{n}_{-\vk,s}}  = \frac{\Big(\widetilde{b}^\dagger_{\vec{k},s}\Big)^{n_{\vk,s}}}{\sqrt{n_{\vk,s} !}}~
\frac{\Big(\widetilde{d}^\dagger_{-\vec{k},s}\Big)^{n_{\vk,s}}}{\sqrt{n_{\vk,s} !}} \ket{0_O} ~~;~~ n_{\vk,s} = 0, 1\,. \label{nouts}\ee where the ``out'' vacuum state $\ket{0_O}$ is such that
\be \widetilde{b}_{\vec{k},s}\ket{0_O}=0 ~~;~~ \widetilde{d}_{\vec{k},s}\ket{0_O}=0\,\,\forall \vk \,, \label{outpairs}  \ee and from eqn. (\ref{nkfer})
\be |B_{k,s}|^2 = \sin^2(\theta_k)= N_k \,.  \label{fersin}\ee
The Schroedinger picture density matrix $\rho_S(\eta) = U_0(\eta,\eta_i) \ket{0_I}\bra{0_I} U^{-1}_0(\eta,\eta_i)$ is now given by
\be \rho_S(\eta) = \Pi_{\vk,s} \Pi_{\vp,s'}  \sum_{n_{\vk,s}=0}^1  \sum_{m_{\vp,s'}=0}^1 \mathcal{C}^*_{m_{\vp,s'}}(p)~ \mathcal{C}_{n_{\vk,s}}(k)~\ket{n_{\vk,s};\overline{n}_{-\vk,s}}\bra{m_{\vp,s'};\overline{m}_{-\vp,s'}}~
e^{2i \int^\eta_{\eta_i} \Big[m_{\vp,s'}\,\omega_p(\eta')-n_{\vk,s}\,\omega_k(\eta') \Big]\,d\eta'} \,, \label{rhosoutfer}\ee where in the fermion case (see appendix (\ref{app:bogofer}))
\be \mathcal{C}_{n_{\vk,s}}(k)  = \cos(\theta_k)\,\Big(-e^{2i\varphi_-(k)}\,\tan(\theta_k) \Bigg)^{n_{\vk,s}} ~~;~~ n_{\vk,s}=0,1 \,.  \label{coefsfer}\ee Just as in the scalar case, the rapid oscillatory phases in the terms that are off-diagonal in pair number $m \neq n$, momenta and helicity average out on time scales $\simeq 1/m \ll 1/H(t)$ leading to the decoherence of the density matrix in this basis. Proceeding as in the scalar case we average these terms over time scales intermediate between $1/m$ and the Hubble time scale $1/H(t)$. This averaging, a coarse graining on the short time scale, is a direct consequence of the separation of time scales during the adiabatic regime, with $H(t)/m \ll 1$ and   yields a density matrix that is diagonal in the basis of particle-antiparticle pairs (\ref{nouts}). The    loss of coherence in the averaging of  correlations implies a loss of information (from these correlations). The calculation of the entropy associated with this loss of information follows the same route as in the scalar case with few modifications consequence of the different statistics. Upon averaging the rapidly varying phases,  the  density matrix becomes diagonal in the basis of particle antiparticle pairs, and is given by
\be \rho^{(d)}_S =  \Pi_{\vk,s} \big[\cos^2(\theta_k) \big]  \sum_{n_{\vk,s}=0}^1 \Big(\tan^2(\theta_k)\Big)^{n_{\vk,s}}
\ket{n_{\vk,s};\overline{n}_{-\vk,s}}\bra{n_{\vk,s};\overline{n}_{-\vk,s}}\,. \label{diagrhofer}\ee We can compare this density matrix with the reduced one obtained by tracing over the antiparticle states,
\be \rho^{(r)}_S(\eta) = \mathrm{Tr}_{\overline{p}}\,\rho_S(\eta) =  \Pi_{\vk,s} \big[\cos^2(\theta_k) \big]  \sum_{n_{\vk,s}=0}^1 \Big(\tan^2(\theta_k)\Big)^{n_{\vk,s}}\ket{n_{\vk,s}}\bra{n_{\vk,s}} \,, \label{reducedrhofer}\ee exhibiting the equivalence of the diagonal matrix elements, namely the probabilities. The density matrices $\rho^{(d)}_S;\rho^{(r)}_S$ feature the \emph{same eigenvalues}, hence the same entropy. Again, this is the statement that the entropy arising from the loss of information in the time averaging or coarse graining, is identical to the entanglement entropy obtained from the reduced density matrix.

The diagonal density matrix (\ref{diagrhofer}) can be written in a familiar quantum statistical mechanics form by introducing a fiducial Hamiltonian
\be \widehat{\mathcal{H}} = \sum_{\vk,s} \mathcal{E}_k\, \widehat{\mathcal{N}}_{\vk,s} \,,\label{fiduHfer} \ee with
\be \mathcal{E}_k = -\ln[\tan^2(\theta_k)]~~;~~ \widehat{\mathcal{N}}_{\vk,s} = \sum_{n_{\vk,s}=0}^1 n_{\vk,s}\,\ket{n_{\vk,s};\overline{n}_{-\vk,s}}\bra{n_{\vk,s};\overline{n}_{-\vk,s}} \,, \label{ENefer}\ee   and the  partition function is given by
\be \mathcal{Z} = \Pi_{\vk,s}[\cos^2(\theta_k)]^{-1} = \Pi_{\vk,s}[1+\tan^2(\theta_k)] \,, \label{zetfer}\ee  so that
\be \rho^{(d)}_S = \frac{e^{-\widehat{\mathcal{H}}}}{\mathcal{Z}}~~;~~ \mathcal{Z}= \mathrm{Tr}\,e^{-\widehat{\mathcal{H}}} \equiv e^{-\mathbb{F}} \,,\label{rhofidufer}\ee with $\mathbb{F}$ the fiducial free energy. We note that in the fermionic case $\widehat{\mathcal{N}}^2_{\vk,s}=\widehat{\mathcal{N}}_{\vk,s}$ therefore for   fixed $\vk,s$ its eigenvalues are $0,1$ and from the relations (\ref{nkfer}, \ref{fersin}) it follows that
\be \tan^2(\theta_k) = \frac{N_k}{1-N_k} \,. \label{tan2fer}\ee

The entropy is now obtained from (\ref{thermorel}) but now  with
\be U = \mathrm{Tr}\,\rho^{(d)}\, \mathcal{H} = \sum_{\vk,s}\frac{\mathcal{E}_k}{e^{\mathcal{E}_k}+1} = \sum_{\vk,s}N_k\,\ln\Big[ \frac{1-N_k}{N_k}\Big] \,.\label{Ufer}\ee The entropy is now given by
\be   S^{(d)} = -2\,\sum_{\vk} \Bigg\{(1-N_k)\,\ln(1-N_k) + N_k\,\ln N_k   \Bigg\}\,.  \label{vNSfer}\ee This is a remarkable result, the entanglement entropy is proportional to  the quantum kinetic entropy for fermions in terms of the distribution function\cite{bernstein}.  The factor $2$ accounts for two helicity eigenstates, since the distribution function is the same for both helicities. We highlight that although the number of particles and of antiparticles are the same, the entropy does \emph{not} feature a factor $4$ (particle, anti-particle with two helicities) but a factor $2$. The reason behind this is the same as in the complex scalar case:  particle and antiparticles are produced in \emph{correlated pairs} not independently. This important aspect  is also at the heart of the equivalence between the entropy arising from dephasing and decoherence and the entanglement entropy: tracing over one member of the particle-anti-particle pairs in (\ref{rhosoutfer}) (either particle or anti-particle) reduces the full density matrix (\ref{rhosoutfer}) to   (for example tracing over anti-particles)
\be \rho^{(r)}(\eta) = \Pi_{\vk,s} \big[\cos^2(\theta_k) \big]  \sum_{n_{\vk,s}=0}^1 \Big(\tan^2(\theta_k)\Big)^{n_{\vk,s}}
\ket{n_{\vk,s}}\bra{n_{\vk,s}}\,, \label{redrhofer}\ee  yielding an entanglement  entropy equivalent to  (\ref{vNSfer}). We also find
\bea && \mathrm{Tr}\,\widetilde{b}^\dagger_{\vec{k},s}\widetilde{b}_{\vec{k},s}\,  \rho^{(d)}_S    =   \mathrm{Tr}\,\widetilde{d}^\dagger_{\vec{k},s}\widetilde{d}_{\vec{k},s}\,  \rho^{(d)}_S = |B_{k,s}|^2= N_k \nonumber \\ && \mathrm{Tr}\,\widetilde{b}^\dagger_{\vec{k},s}\widetilde{d}^\dagger_{-\vec{k},s}\,  \rho^{(d)}_S    = \mathrm{Tr}\,\widetilde{d}_{-\vec{k},s}\widetilde{b}_{\vec{k},s}\,  \rho^{(d)}_S = 0 \,.\label{idsdiag}  \eea Therefore, the   energy density  and pressure near matter radiation equality when the adiabatic approximation is very reliable and the density matrix has undergone complete decoherence via dephasing,   are given by
\be \overline{\rho}(\eta) = \mathrm{Tr}\widehat{\overline{\rho}}(\eta)\, \rho^{(d)}_S ~~;~~\overline{P}(\eta) = \mathrm{Tr}\widehat{\overline{P}}(\eta)\, \rho^{(d)}_S\,, \label{efpf}\ee
these are obtained
to leading (zeroth) order in the adiabatic approximation by using the spinors   (\ref{Uspinrdasy},\ref{Vspinrdasy}). As a consequence of decoherence yielding the identities (\ref{idsdiag}), the particle-antiparticle interference terms vanish. Because the  spinors (\ref{Uspinrdasy},\ref{Vspinrdasy}) are eigenstates of the instantaneous conformal Hamiltonian (\ref{enesig}) with eigenvalues $\pm \omega_k(\eta)$,  we find to leading order in the adiabatic expansion\footnote{For higher order contributions see ref.\cite{herringfer}.}
     \be \overline{\rho}(\eta) =  \underbrace{-\frac{1}{\pi^2 a^4(\eta)} ~ \int^\infty_0 k^2 dk\,\omega_k(\eta)
}_{\overline{\rho}_0(\eta)}+ \underbrace{\frac{2}{\pi^2 a^4(\eta)} ~ \int^\infty_0 k^2 dk\,N_k~\omega_k(\eta)}_{\overline{\rho}_{p\overline{p}}(\eta)}\,,\label{rhomatasy}  \ee

\be \overline{P}(\eta) = \underbrace{-\frac{1}{3\pi^2 a^4(\eta)} ~ \int^\infty_0 k^2 dk\,\frac{k^2}{\omega_k(\eta)}
}_{\overline{P}_0(\eta)}+ \underbrace{\frac{2}{3\pi^2 a^4(\eta)} ~ \int^\infty_0 k^2 dk\,N_k~\frac{k^2}{\omega_k(\eta)}}_{\overline{P}_{p\overline{p}}(\eta)}\,,\label{Pmatasy}\ee
where $\overline{\rho}_0(\eta),\overline{P}_0(\eta)$ are the zero point energy density and pressure and $\overline{\rho}_{p\overline{p}}(\eta),\overline{P}_{p\overline{p}}(\eta)$ are the contributions from gravitational particle production. The zero point and particle production contributions independently obey covariant conservation. As explained in ref.\cite{herringfer} the zero point contribution is absorbed into a renormalization\cite{rio,ferreiro,barbero,ghosh,landete}, therefore the kinetic-fluid description of gravitationally produced fermionic dark matter near matter radiation equality can now be summarized as

\be \mathcal{N}_{p\overline{p}} =  \frac{2}{\pi^2} ~ \int^\infty_0 k^2 \,N_k  \,dk\,,\label{Nppfer}  \ee

\be \overline{\rho}_{p\overline{p}}(\eta) =     \frac{2}{\pi^2 a^4(\eta)} ~ \int^\infty_0 k^2 \,N_k~\omega_k(\eta) \,dk\,,\label{rhoppfer}  \ee

\be \overline{P}_{p\overline{p}}(\eta) =    \frac{2}{3\pi^2 a^4(\eta)} ~ \int^\infty_0 k^2 \,N_k~\frac{k^2}{\omega_k(\eta)} \, dk\,,\label{Pppfer}\ee

\be \mathcal{S}_{p\overline{p}} = -\frac{2}{2 \pi^2} \int^\infty_0 k^2\,\Big\{(1-N_k)\,\ln(1-N_k) + N_k\,\ln N_k   \Big\}  dk \,,\label{Sfer}\ee where $\mathcal{N}_{p\overline{p}}$ is the total comoving number density of particles plus antiparticles  produced, $\mathcal{S}_{p\overline{p}}$ is the time independent comoving entropy density, and  the distribution function $N_k$ is given by eqn. (\ref{almMB}).  The kinetic fluid forms of the energy density (\ref{rhoppfer}) and pressure (\ref{Pppfer}) are exactly the same as obtained in ref.\cite{herringfer} by averaging over the fast phases in the particle-antiparticle interference terms.
 Therefore, just as in the bosonic case this averaging in the energy momentum tensor and the emergence of the kinetic fluid form in the adiabatic regime is a direct manifestation of  decoherence by dephasing in the density matrix, hence also directly related to the emergence of entropy in this case.

With the distribution function (\ref{almMB}), we find
\be \mathcal{N}_{p\overline{p}} = \frac{2}{\pi^2}\, \Big(2mT_H \Big)^{3/2}\,\times 0.126 \,, \label{numint}\ee  and
\be  \mathcal{S}_{p\overline{p}} =  \frac{1}{\pi^2}\, \Big(2mT_H \Big)^{3/2}\,\times 0.451 \,, \label{Sint} \ee with a specific entropy
\be \frac{\mathcal{S}_{p\overline{p}}}{\mathcal{N}_{p\overline{p}}}\simeq 1.8 \,. \label{entrat}\ee We note that a specific entropy $\mathcal{O}(1)$ is typical of a  thermal species.
However, with $m\simeq 10^{8}\,\mathrm{GeV}$ for a heavy fermion with the correct dark matter abundance\cite{herringfer}, the ratio of its comoving entropy to that of the (CMB) today given by (\ref{encmb}) which also features a specific entropy $\mathcal{O}(1)$,  is
\be \frac{\mathcal{S}_{p\overline{p}}}{\mathcal{S}_{cmb}} \simeq 10^{-15}\,, \label{ratcmbfer}\ee therefore even for a heavy fermionic dark matter species that is gravitationally produced, its entropy is negligible compared to that of the (CMB) today.

\section{Discussion}\label{sec:discussion}

\textbf{Real scalars, Majorana fermions:} We have studied complex scalars and Dirac fermions for which particles are different from antiparticles. However, the results apply just as well to real scalars and Majorana fermions, in which cases particles are the same as antiparticles and the correlated pair states are now of the form $\ket{n_{\vk},n_{-\vk}}$. The entanglement entropy is exactly the same as for complex scalars or Dirac fermions respectively, since for each value of $\vk$ (and helicity $s$ for fermions), tracing over one member of the pair (say that with $-\vk$ ) yields exactly the same probabilities, regardless of whether it is a particle or an antiparticle. This is also explicit in the entanglement entropies obtained above since there is no factor $2$ for particle and antiparticle, because of the correlated nature of the pair state, independently of whether the members of the pairs are particle and antiparticle or particle-particle with opposite momenta.

\textbf{The origin of entropy: the ``out'' basis is a pointer basis.} In the language of quantum information, the ``out'' basis of particles is the ``measured'' basis and constitutes a \emph{pointer basis}\cite{zurek}. This is  indeed a privileged basis, since the energy momentum tensor in this out particle basis describes the abundance, equation of state and entropy of \emph{particles} (and antiparticles). These are the observable macroscopic variables that describe the properties of dark matter. It is precisely in this basis that the rapid dephasing and coarse graining as a consequence of time averaging over the short time scales leads to decoherence and information loss, with the concomitant emergence of a non-vanishing entropy.

 One could   take  expectation values of the energy momentum tensor  (or any other observable related to dark matter) in the ``in'' vacuum state $\ket{0_I}$ or the density matrix $\ket{0_I}\bra{0_I}$ as is the case in refs.\cite{herring,herringfer}. This expectation value features the rapidly oscillating interference terms between ``out'' particles and antiparticles, which were averaged out on the short time scales in these references. This averaging in the expectation values in the ``in'' state $\ket{0_I}$  are a manifestation of the loss of correlations by dephasing, yet do not make explicit the \emph{entropic} content of this decoherence process.

These are precisely the coherences and correlations that are averaged out in the density matrix in the Schroedinger picture in the out basis. Hence, particle ``observables'' or measurements in the out particle basis in general will undergo this process of decoherence via dephasing even when  the matrix elements are obtained in the ``in'' basis. The coarse graining of the density matrix in the Schroedinger picture in the out basis  exhibits directly this     decoherence mechanism by dephasing and the emergence of entropy. It also makes explicit that the decoherence time scale is $\simeq 1/m$. Therefore, the origin of entropy is deeply associated with this natural selection of basis of ``out particles'' to describe the density matrix and the statistical  properties of dark matter.

\textbf{More general arguments for entropy:} Although we focused on the entropy in gravitational particle production, the main concepts elaborated here are more general. For example they apply also to the case when particles are produced from inflaton oscillations at the end of inflation\cite{vela},  or by parametric resonance during reheating\cite{reheat1,reheat}. In these cases, a homogeneous scalar field (generically the inflaton) couples non-linearly to the matter bosonic or fermionic fields. If the expectation value of this scalar field depends on time, acting as a time dependent mass term, such coupling leads to production of particle or particle-antiparticle pairs entangled in momentum (and any other conserved quantum number). The ``in'' basis is  generically a superposition of the out particle basis states, therefore the interference effects will also be manifest in a similar manner as studied here, although the occupation number of ``out'' states will be different for different mechanisms. Because dark matter   particles are defined as asymptotic out states in the adiabatic era, a separation of time scales as in the  adiabatic Schroedinger picture in which the density matrix   evolves in time will feature a structure very similar to that unveiled in the study above, but with different   probabilities determined by the different processes. Nevertheless dephasing and decoherence will play a similar role leading to an entropy of the very same form as obtained above but with different $N_k$.

\textbf{Entanglement entropy vs. entropy (isocurvature) perturbations:} The entanglement entropy discussed above should not be identified with linear entropy or isocurvature \emph{perturbations}. The latter are generically associated with multiple fields with non-vanishing expectation values during inflation\cite{gordon,byrnes,bartolo}.   Entropy perturbations in the case when scalar fields do \emph{not} acquire expectation values\cite{sena}, or for fermionic fields (which cannot acquire expectation values) \cite{chungiso} were analyzed within the context of zero point contributions to the energy momentum tensor in refs.\cite{sena,chungiso}. However, in refs.\cite{herring,herringfer} it was argued that the   renormalization fully subtracting the zero point contribution as is implicitly or explicitly done in the literature, prevents a consistent interpretation of entropy perturbations from  the zero point contribution of the energy momentum tensor as advocated in refs.\cite{sena,chungiso}. In our study here the scalar field does \emph{not acquire} an expectation value  and we implemented the same renormalization scheme subtracting completely the zero point contribution to the energy momentum tensor as in refs. \cite{herring,herringfer} both for scalar and fermion fields. Therefore   the analysis and conclusions of refs.\cite{sena,chungiso}   do not apply to our study.

Curvature perturbations and inhomogeneous gravitational potentials will modify the entropies (\ref{spp},\ref{vNSfer}) by modifying the distribution functions $N_k \rightarrow N_k + \delta N_k(\vx,t)$ thereby inducing a perturbation in the entanglement entropy. Such perturbation is completely determined by the change in the distribution function which obeys a   linearized collisionless Boltzmann equation in presence of the metric perturbations. This equation along with a proper assessment of initial conditions must be studied in detail for a definite understanding of entropy perturbations, a task that is well beyond the scope and objective of our study.

\vspace{2mm}

\section{Conclusions and further questions:}\label{sec:conclusions}

While the evidence for dark matter is overwhelming, direct detection of a particle physics candidate with interactions with (SM) degrees of freedom, necessary for detection, has proven elusive. Therefore dark matter particles  featuring only gravitational interaction are logically a suitable alternative. Such candidates are produced gravitationally via cosmological expansion,  a phenomenon that received substantial attention in the last few years.
In this article we studied the emergence of entropy in gravitational production of dark matter particles, focusing on the cases of a complex scalar and a Dirac fermion under a minimal set of assumptions as in refs.\cite{herring,herringfer}. We considered a rapid transition from inflation to radiation domination and focused on comoving super-Hubble wavelengths at the end of inflation, with dark matter fields being in their Bunch-Davies vacua during inflation. The ``out'' states are correlated particle-antiparticle pairs and the distribution function of gravitationally produced particles is obtained exactly both for ultra-light scalars and heavier fermions.

Well after the transition and before matter radiation equality there ensues a period of adiabatic evolution when the scale factor $a_{eq} \gg a(t) \gg 10^{-17}/\sqrt{m(eV)}$ characterized by the adiabatic ratio $H(t)/m \ll 1$ with $H(t)$ the Hubble expansion rate and $m$ the particle's mass. During this regime there is a wide separation of time scales with  $1/H(t)$ a long time scale of  cosmological evolution and $1/m$ a short time scale associated with particle dynamics.  As shown in refs.\cite{herring,herringfer}, during this regime
the energy momentum tensor written in the ``out'' particle basis (dark matter particles) feature rapidly varying particle-antiparticle interference terms. Averaging these contributions on intermediate time scales renders the energy momentum tensor of the usual kinetic fluid form.  We show that these rapidly varying interference terms are manifest in the density matrix in the adiabatic Schroedinger picture in the out particle basis  as off diagonal density matrix elements that feature rapid dephasing on short decoherence time scales $\simeq 1/m$. Decoherence by dephasing  effectively reduces the density matrix to a diagonal form in the out basis with a non-vanishing von Neumann entropy. In turn, the von Neumann entropy is exactly the same as the \emph{entanglement entropy} obtained by tracing over one member of the correlated particle-antiparticle pair.

Remarkably, we find that the comoving von-Neumann-entanglement entropy density is \emph{almost} of the  kinetic fluid form in terms of the distribution function $N_k$
\be \mathcal{S}_{p\overline{p}} = \pm \frac{1}{2 \pi^2} \int^\infty_0 k^2\,\Big\{(1\pm N_k)\,\ln(1 \pm N_k) \mp  N_k\,\ln N_k   \Big\}  dk \,,\label{SvNen} \ee where $(+)$ is for \emph{real or complex}  bosons and $(-)$ is for each spin/helicity of \emph{Dirac or Majorana} fermions. If the ``out'' states were described by \emph{independent} particles and/or antiparticles, complex bosons and Dirac fermions would have twice the number of degrees of freedom of real bosons and Majorana fermions and the entropy would feature an extra factor $2$ when particles are different from antiparticles.  The fact that the entanglement entropies are the same regardless of whether particles are different from antiparticles   is a consequence of the pair correlations of the ``out'' state, explaining the qualifier  ``\emph{almost}''.  These particle-antiparticle or particle-particle pairs are entangled in momentum (and helicity in the case of fermions) and the entanglement entropy, obtained by tracing over one member of the pair is the \emph{same} in both cases regardless of whether  particles are the same or different from antiparticles. An important conclusion of our study  is that  the  von Neumann-entanglement- entropy and the kinetic fluid form of the energy momentum are all a consequence of decoherence of the density matrix in the out basis.

We argue that the origin of entropy is deeply related to the natural physical basis of ``out'' particles that determine the statistical properties of dark matter, such as energy density, pressure and entropy. Furthermore, we also argue that our results are more general and apply also to several other production mechanisms such as parametric amplification and production from inflaton oscillations at the end of inflation.

For an ultra-light bosonic dark matter candidate minimally coupled to gravity we find that  while the occupation number is very large in the infrared region,  the specific entropy, or entropy per particle, is negligibly small, indicating that this dark matter candidate is produced in a \emph{condensed state}, albeit with vanishing expectation value. For     fermionic dark matter the distribution function is nearly thermal\cite{herringfer} and   the specific entropy is $\mathcal{O}(1)$ consistent with a thermal species.

\vspace{2mm}

\textbf{Further questions:}

\textbf{a) Observational consequences?:} While the energy density and pressure (or equation of state)  both have clear observational consequences and directly yield information on clustering properties such as the free streaming length or cut-off in the matter power spectrum\cite{herring},  we have not yet identified an observational consequence directly associated with entropy. As discussed above, for both cases, ultra light or heavier fermionic gravitationally produced dark matter, their comoving entropy is many orders of magnitude smaller than that for the (CMB) today.

The similarity with the fluid kinetic form suggests that perhaps the entropy \emph{may} play a role in the dynamics of galaxy formation. Pioneering work in refs.\cite{lb,tremaine} studied the non-equilibrium process  of violent relaxation in collisionless galactic dynamics in terms of an H-function that is similar to the statistical entropy of  a classical dilute gas. It is argued in these references that such H-function increases during this process of relaxation towards an equilibrium state. It is an intriguing possibility that the entanglement entropy that we find \emph{could} play a similar role in understanding the evolution of clustering during the matter dominated era.

Another important question is the role of metric perturbations on the entropy, as mentioned above this would entail a study of the linearized boltzmann equation and further understanding on initial conditions.

\vspace{2mm}

\textbf{b) Interactions:}

Although we did not consider the possibility of dark matter self-interactions or interactions with (SM) degrees of freedom, the study of how the entanglement entropy evolves in time as a consequence of such interactions would be of fundamental interest and a worthy endeavor. In principle the evolution of the entropy could be obtained by setting up a quantum kinetic Boltzmann  equation for the distribution function $N_k$. However, a new framework must be developed to implement this program, because typically the Boltzmann equation is obtained by calculating transition amplitudes in S-matrix theory, however the mode functions even during the adiabatic regime are \emph{not} the same as in Minkowski space time. Furthermore, the usual approach takes the infinite time limit to obtain the transition probabilities, which   in principle is  not warranted   in presence of cosmological expansion, instead a framework similar to that implemented in refs.\cite{decay1,decay2} must be adapted to a quantum kinetic approach.

The first law of thermodyamics when combined with covariant conservation of the energy entails that the total \emph{thermodynamic} entropy is constant, namely the cosmological expansion is adiabatic in the thermodynamic sense in agreement with the Universe being a closed system. However, the entanglement entropy is \emph{not} a thermodynamic entropy, therefore  if interactions are included, it is by no means clear that  that the entanglement entropy remains constant. Ref.\cite{kandrup} advocated  a possible statistical framework to include interactions akin to the  Bogoliubov-Born-Green-Kirkwood-
Yvon (BBGKY) hierarchy of equations that yields the usual Boltzmann equation. While this suggestion is compelling, the applicability of such framework to study the time evolution of the entanglement entropy  merits further study beyond the scope of this article.

\appendix

\section{Bogoliubov  Transformation for Bosonic fields}\label{app:bogobose}

The unitary operator that implements the Bogoliubov transformation (\ref{bops})
\bea
c_{\vk} &  =  & a_{\vk}\,A_{k}+b_{-\vk}^{\dagger}\,B_{k}^{*}\,,\\
d^\dagger_{-\vk} & = & b^\dagger_{-\vk}\,A^*_{k}+a_{\vk} \,B_{k}  \, , \label{bogotra}
\eea
 is obtained as follows. The coefficients $A_k~;~B_k$ are functions solely of $k$ determined by the relations (\ref{ABcoefs}) and obey the condition (\ref{condiAB}). We write
 \be A_k = e^{i\varphi_A(k)}\,\cosh(\theta_k)~~;~~B_k = e^{i\varphi_B(k)} \,\sinh(\theta_k) \,. \label{ABdefs} \ee
 Let us introduce the following definitions (we suppress the momentum arguments of the angles):
 \bea && \varphi_A = \varphi_+ + \varphi_-~~;~~ \varphi_B = \varphi_+ - \varphi_- \nonumber \\
&&  a_{\vk}\,e^{i\varphi_+} = \widetilde{a}_{\vk}~~;~~  b_{\vk}\,e^{i\varphi_+} = \widetilde{b}_{\vk}\nonumber \\
&& c_{\vk}\,e^{-i\varphi_-} = \widetilde{c}_{\vk} ~~;~~ d_{\vk}\,e^{-i\varphi_-} = \widetilde{d}_{\vk} \label{newdefs}\,,\eea in terms of which the transformation (\ref{bogotra}) becomes
\bea \widetilde{c}_{\vk} & = & \widetilde{a}_{\vk}\,\cosh(\theta_k) + \widetilde{b}_{-\vk}\,\sinh(\theta_k) \label{trafo1}\\
\widetilde{d}^\dagger_{-\vk} & = & \widetilde{b}^\dagger_{-\vk}\,\cosh(\theta_k) + \widetilde{a}_{\vk}\,\sinh(\theta_k) \,. \label{trafo2}\eea These transformations are implemented by the following unitary operator
\be S[\theta] = \Pi_{\vk}\,\exp\Big\{\theta_k \,\Big[\widetilde{b}_{-\vk}\,\widetilde{a}_{\vk}- \widetilde{a}^\dagger_{\vk}\,\widetilde{b}^\dagger_{-\vk} \Big]\Big\}~~;~~ S^{-1}[\theta] = S[-\theta] \,, \label{Strfo}\ee so that

\bea S[\theta]\, \widetilde{a}_{\vk}\,S^{-1} [\theta] & = &  \widetilde{c}_{\vk} \label{bogoc}\\
 S[\theta]\, \widetilde{b}^\dagger_{-\vk}  \,S^{-1} [\theta] & = &  \widetilde{d}_{-\vk}\,, \label{bogod}\eea as can be confirmed by expanding the exponential and using the canonical commutation relations. An important identity yields the following factorization of the exponential\cite{barnett},
 \bea S[\theta] & = &  \Pi_{\vk}\,\exp\Big\{-\ln(\cosh(\theta_k)) \Big\}~  \exp\Big\{-\tanh(\theta_k)\,\widetilde{a}^\dagger_{\vk}\,\widetilde{b}^\dagger_{-\vk}   \Big\}~ \exp\Big\{-\ln(\cosh(\theta_k)\,\Big(\widetilde{a}^\dagger_{\vk}\,\widetilde{a}_{\vk}+ \widetilde{b}^\dagger_{\vk}\,\widetilde{b}_{\vk} \Big)  \Big\}\nonumber \\
 & \times &  \exp\Big\{\tanh(\theta_k)\,\widetilde{b}_{-\vk}\,\widetilde{a}_{\vk}   \Big\} \label{factorS}  \,.  \eea

 The inverse Bogoliubov transformation is given by
 \bea  \widetilde{a}_{\vk} & = &  \widetilde{c}_{\vk}\,\cosh(\theta_k) - \widetilde{d}^\dagger_{-\vk}\,\sinh(\theta_k) \nonumber \\
 \widetilde{b}^\dagger_{-\vk} & = & \widetilde{d}^\dagger_{-\vk}\,\cosh(\theta_k)- \widetilde{c}_{\vk}\,\sinh(\theta_k)\,.  \label{invbogo}\eea The unitary operator that implements it
 is
 \be T[\theta] = \Pi_{\vk}\,\exp\Big\{-\theta_k \,\Big[\widetilde{c}_{\vk}\,\widetilde{d}_{-\vk} - \widetilde{d}^\dagger_{-\vk}\,\widetilde{c}^\dagger_{\vk} \Big]\Big\}~~;~~ T^{-1}[\theta] = T[-\theta] \,, \label{Ttrfo}\ee so that

 \bea T[\theta]\, \widetilde{c}_{\vk}\,T^{-1} [\theta] & = & \widetilde{a}_{\vk} \nonumber \\
 T[\theta]\, \widetilde{d}^\dagger_{-\vk}\,T^{-1} [\theta] & = & \widetilde{b}^\dagger_{-\vk}\,. \label{Tbog}\eea
 The factorized form of $T[\theta]$ is
 \bea T[\theta] & = &  \Pi_{\vk}\,\exp\Big\{-\ln(\cosh(\theta_k)) \Big\}~  \exp\Big\{\tanh(\theta_k)\,\widetilde{c}^\dagger_{\vk}\,\widetilde{d}^\dagger_{-\vk}   \Big\}~ \exp\Big\{-\ln(\cosh(\theta_k)\,\Big(\widetilde{c}^\dagger_{\vk}\,\widetilde{c}_{\vk}+ \widetilde{d}^\dagger_{\vk}\,\widetilde{d}_{\vk} \Big)  \Big\}\nonumber \\
 & \times &  \exp\Big\{-\tanh(\theta_k)\,\widetilde{d}_{-\vk}\,\widetilde{c}_{\vk}   \Big\} \label{factorT}  \,.  \eea

 These operators allow us to relate the ``in'' vacuum state to ``out'' states. Define the ``out'' vacuum state $\ket{0_O}$ as that annihilated by $c_{\vk};d_{\vk}$, namely
 \be c_{\vk}\,\ket{0_O} =0 ~~;~~ d_{\vk}\,\ket{0_O} =0\,.  \label{outvac}\ee Pre-multiplying these expressions by $T[\theta]$ and inserting $T^{-1}[\theta]\,T[\theta]=1$,  yields
 \be \underbrace{\Big( T[\theta]\,c_{\vk}\,T^{-1}[\theta]\Big)}_{a_{\vk}}\,\underbrace{\Big(T[\theta]\,\ket{0_O}\Big)}_{\ket{0_I}} =0 ~~;~~\underbrace{ \Big( T[\theta]\,d_{\vk}\,T^{-1}[\theta]\Big)}_{b_{\vk}}\,\underbrace{\Big(T[\theta]\,\ket{0_O}\Big)}_{\ket{0_I}} =0\,.   \label{Toutvac}\ee

Therefore, we find
\be \ket{0_I} = \Pi_{\vk}\Bigg\{\Big[\cosh(\theta_k)\Big]^{-1}~ \,\sum_{n_{\vk}=0}^\infty \Bigg(e^{2i\varphi_-(k)}\,\tanh(\theta_k) \Bigg)^{n_{\vk}} \ket{n_{\vk};\overline{n}_{-\vk}}\Bigg\} \,, \label{vacin}\ee where the ``out'' particle-antiparticle states
\be  \ket{n_{\vk};\overline{n}_{-\vk}}  = \frac{\Big(c^{\dagger}_{\vk}\Big)^{n_{\vk}}}{\sqrt{n_{\vk} !}}~
\frac{\Big(d^{\dagger}_{-\vk}\Big)^{n_{\vk}}}{\sqrt{n_{\vk} !}} \ket{0_O} ~~;~~ n_{\vk} = 0,1,2 \cdots\,. \label{noutsbose}\ee

In quantum optics these correlated states are known as two-mode squeezed states\cite{barnett}.  Several checks are in order:
\be \langle 0_I|0_I\rangle = \Pi_{\vk} \frac{1}{\cosh^2(\theta_k)}\, \sum^{\infty}_{n=0}(\tanh^2(\theta_k))^n = \Pi_{\vk}  \frac{1}{\cosh^2(\theta_k)}\,\frac{1}{1-\tanh^2(\theta_k)} = 1 \,, \label{ck1}\ee

\be \bra{0_I}c^\dagger_{\vp}c_{\vp}\ket{0_I} =  \bra{0_I}d^\dagger_{\vp}d_{\vp}\ket{0_I} =    \frac{1}{\cosh^2(\theta_p)}\, \sum^{\infty}_{n=0}n\, (\tanh^2(\theta_p))^n = \sinh^2(\theta_p) = |B_p|^2 \,, \label{ck2}\ee

\bea \bra{0_I}c^\dagger_{\vp}d^\dagger_{\vp}\ket{0_I}  & = &       \frac{1}{\cosh^2(\theta_p)}\, \frac{e^{-2i\varphi_-(p)}}{\tanh(\theta_p)} \sum^{\infty}_{n=0}(1+n)\, (\tanh^2(\theta_p))^{1+n} = \frac{e^{-2i\varphi_-(p)}}{\tanh(\theta_p)}\,\frac{\tanh^2(\theta_p)}{\cosh^2(\theta_p)}\,
\frac{1}{\Big(1-\tanh^2(\theta_p)\Big)^2} \nonumber \\ &  =  &  e^{-2i\varphi_-(p)}\,\sinh(\theta_p)\,\cosh(\theta_p) = B_p\,A^*_p   \,, \label{ck3}\eea  thereby confirming the identities (\ref{parts}) in the ``out'' basis.

\section{Bogoliubov transformation for Fermionic fields}\label{app:bogofer}

The Bogoliubov transformations for fermionic operators are somewhat more subtle because of the anticommutation relations.
The out basis operators are related to the in basis via the Bogoliubov transformation
\bea \widetilde{b}_{\vec{k},s} & = &   {b}_{\vec{k},s} A_k - {d}^{\dagger}_{-\vec{k},s} B^*_{k,s} \label{btilap} \\
\widetilde{d}^{\,\dagger}_{-\vec{k},s} & = &   {d}^{\dagger}_{-\vec{k},s} A^*_{k,s} + {b}_{\vec{k},s} B_{k,s} \,, \label{ddtilap} \eea  and
\be |A_{k,s}|^2 + |B_{k,s}|^2 =1 \,. \ee
We write
\be A_{k,s} = \cos(\theta_k) \,e^{i(\varphi_+ + \varphi_-)}~~;~~ B_{k,s} = \sin(\theta_k) \,e^{i(\varphi_+ - \varphi_-)} \label{ABdefsfer}\ee where the $k,s$  arguments of the phases are implicit. We now absorb the phases into a redefinition of the various operators,
\bea \widetilde{b}_{\vec{k},s} \equiv \widetilde{b}_{\vec{k},s}\,e^{-i\varphi_-} ~~;~~ \widetilde{d}^{\,\dagger}_{-\vec{k},s} \equiv \widetilde{d}^{\,\dagger}_{-\vec{k},s}\,e^{i\varphi_-} \nonumber \\ {b}_{\vec{k},s} \equiv {b}_{\vec{k},s}\,e^{i\varphi_+}~~;~~ {d}^{\,\dagger}_{-\vec{k},s} \equiv {d}^{\,\dagger}_{-\vec{k},s}\,e^{-i\varphi_+}\,.  \label{bdphases}\eea In terms of these redefinitions the Bogoliubov transformations (\ref{btilap},\ref{ddtilap}) read
\bea \widetilde{b}_{\vec{k},s} & = &   {b}_{\vec{k},s} \cos(\theta_k) - {d}^{\dagger}_{-\vec{k},s} \sin(\theta_k) \label{btil2} \\
\widetilde{d}^{\,\dagger}_{-\vec{k},s} & = &   {d}^{\dagger}_{-\vec{k},s} \cos(\theta_k) + {b}_{\vec{k},s} \sin(\theta_k) \,. \label{ddtil2} \eea The inverse transformation is
\bea  {b}_{\vec{k},s} & = &   \widetilde{b}_{\vec{k},s} \cos(\theta_k) + \widetilde{d}^{\dagger}_{-\vec{k},s} \sin(\theta_k) \label{btil2inv} \\
 {d}^{\,\dagger}_{-\vec{k},s} & = &   \widetilde{d}^{\dagger}_{-\vec{k},s} \cos(\theta_k) - \widetilde{b}_{\vec{k},s} \sin(\theta_k) \,. \label{ddtil2inv} \eea
  It is convenient to define
\be \gamma_{\vk} = \widetilde{b}^\dagger_{\vec{k},s}\,\widetilde{d}^{\dagger}_{-\vec{k},s} - \widetilde{d}_{-\vec{k},s}\,\widetilde{b}_{\vec{k},s}\,, \label{gam}\ee in terms of which, this inverse transformation is generated by the unitary operator
 \be T_f[\theta_k] = \exp\big\{-\theta_k\,\gamma_{\vk}  \big\} \,,  \label{Sferop}\ee namely
 \bea {b}_{\vec{k},s} & = &  T_f[\theta_k]\,\widetilde{b}_{\vec{k},s}\,T^{-1}_f[\theta_k]\label{bofteta}\\
{d}^{\,\dagger}_{-\vec{k},s} & = &   T_f[\theta_k]\,\widetilde{d}^\dagger_{-\vec{k},s}\,T^{-1}_f[\theta_k]\,. \label{ddofteta}\eea To see that this is the case, consider the definitions
 \bea \alpha(\theta)  & = &  T_f[\theta]\,\widetilde{b}_{\vec{k},s}\,T^{-1}_f[\theta ]\label{alfa}\\
\beta(\theta) & = &   T_f[\theta ]\,\widetilde{d}^\dagger_{-\vec{k},s}\,T^{-1}_f[\theta]\,. \label{beta}\eea Using the anticommutation relations we find
\bea \frac{d\alpha(\theta)}{d\theta} & = & \beta(\theta) \label{dalfa}\\
\frac{d\beta(\theta)}{d\theta} & = & -\alpha(\theta)\,, \label{dbeta}\eea
with the ``initial conditions''
\bea \alpha(0) & = & \widetilde{b}_{\vec{k},s}~~;~~ \frac{d\alpha(\theta)}{d\theta}\Big|_{\theta=0} = \beta(0) = \widetilde{d}^\dagger_{-\vec{k},s} \label{inialfa}\\
\beta(0) & = & \widetilde{d}^\dagger_{-\vec{k},s}~~;~~ \frac{d\beta(\theta)}{d\theta}\Big|_{\theta=0} = -\alpha(0) = -\widetilde{b}_{\vec{k},s} \label{inibeta} \,. \eea The solutions of equations (\ref{dalfa},\ref{dbeta}) with the initial conditions (\ref{inialfa},\ref{inibeta}) are given by
\bea \alpha(\theta)  & = &  \widetilde{b}_{\vec{k},s} \,\cos(\theta) + \widetilde{d}^\dagger_{-\vec{k},s}\,\sin(\theta) \label{alfasolu}\\
\beta(\theta)  & = &  \widetilde{d}^\dagger_{-\vec{k},s} \,\cos(\theta) - \widetilde{b}_{\vec{k},s}\,\sin(\theta)\,,  \label{betasolu} \eea which are recognized as ${b}_{\vec{k},s},{d}^{\,\dagger}_{-\vec{k},s}$ equations (\ref{btil2inv},\ref{ddtil2inv}) respectively, confirming the relations (\ref{bofteta},\ref{ddofteta}).  These relations may also be found from the identity
\be
e^{X}Ye^{-X}=Y+[X,Y]+\frac{1}{2!}[X,[X,Y]]+...
\ee with $X= -\theta_k\,\gamma_{\vk}$ and $Y= \widetilde{b},\widetilde{d}^\dagger$ respectively. Suppressing the indics, $\vk,s$, it follows that
\be
e^{-\theta\gamma}\tilde{b}e^{\theta\gamma}=\tilde{b}+\theta\tilde{d}^{\dagger}-\frac{\theta^{2}}{2!}\tilde{b}-\frac{\theta^{3}}{3!}\tilde{d}^{\dagger}...
\ee

\be
=\tilde{b}(1-\frac{\theta^{2}}{2!}+\frac{\theta^{4}}{4!}...)+\tilde{d}^{\dagger}(\theta-\frac{\theta^{3}}{3!}+....)
\ee
\be
\Rightarrow e^{-\theta\gamma}\tilde{b}e^{\theta\gamma}=\tilde{b}\cos\theta+\tilde{d^{\dagger}}\sin\theta=b \,.
\ee

Similarly,
\be
e^{-\theta\gamma}\tilde{d}^{\dagger}e^{\theta\gamma}=\tilde{d}^{\dagger}-\theta\tilde{b}-\frac{\theta^{2}}{2!}\tilde{d}^{\dagger}+\frac{\theta^{3}}{3!}\tilde{b}...
\ee

\be
\Rightarrow e^{-\theta\gamma}\tilde{d}^{\dagger}e^{\theta\gamma}=\tilde{d}^{\dagger}\cos\theta-\tilde{b}
\sin\theta=d^{\dagger} \,.
\ee

In order to find a more compact expression for $T_f[\theta]$ it proves convenient to expand,

\be T_f[\theta_k] = 1-\theta_k\,\gamma_{\vk} + \frac{1}{2!}\,\theta^2_k\,\gamma^2_{\vk} + \frac{1}{3!}\,\theta^3_k\,\gamma^3_{\vk} + \cdots \label{sexp}\ee Using the canonical anticommutation relations we find
\be \gamma^2_{\vk} = - \Big[ \widetilde{b}^\dagger_{\vec{k},s}\, \widetilde{b}_{\vec{k},s}\, \widetilde{d}^\dagger_{-\vec{k},s}\, \widetilde{d}_{-\vec{k},s}+ \widetilde{d}_{-\vec{k},s}\, \widetilde{d}^\dagger_{-\vec{k},s}\,\widetilde{b}_{\vec{k},s}\, \widetilde{b}^\dagger_{\vec{k},s}  \Big] = - \mathbf{P}_{\vk}\,. \label{gam2}\ee $\mathbf{P}_{\vk}$ is a projection operator, which in terms of \be \widetilde{b}^\dagger_{\vec{k},s}\, \widetilde{b}_{\vec{k},s} = \widehat{n}_{\vk}~~;~~\widetilde{d}^\dagger_{-\vec{k},s}\, \widetilde{d}_{-\vec{k},s} = \widehat{\overline{n}}_{-\vk}\,, \label{ns}\ee
  may also be written as
\be \mathbf{P}_{\vk} = \widehat{n}_{\vk}\,\,\,\widehat{\overline{n}}_{-\vk}+ (1-\widehat{n}_{\vk})\,(1-\widehat{\overline{n}}_{-\vk})~~;~~ \mathbf{P}^2_{\vk} = \mathbf{P}_{\vk}\,. \label{projop}\ee Again using the anticommutation relations we find
\be \gamma_{\vk}\,\mathbf{P}_{\vk} = \mathbf{P}_{\vk}\,\gamma_{\vk} = \gamma_{\vk}\,, \label{itergam}\ee iterating yields
\be \gamma^3_{\vk} = - \gamma_{\vk}~~;~~ \gamma^4_{\vk} = \mathbf{P}_{\vk} ~~;~~ \gamma^5_{\vk} = \gamma_{\vk}\,\mathbf{P}_{\vk} =  \gamma_{\vk} \cdots \label{series}\ee Combining these results we finally find
\be T_f[\theta_k] = 1- \mathbf{P}_{\vk} + \mathbf{P}_{\vk}\,\cos(\theta_k) - \gamma_{\vk}\,\sin(\theta_k)\,. \label{Sffin}\ee Since the operators $\gamma_{\vk}$ commute  for different values of $\vk$ it follows that the full unitary transformation is
\be T_f[\theta] = \Pi_{\vk} T_f[\theta_k]\,.  \label{fullSf}\ee

 Define the ``out'' vacuum state $\ket{0_O}$ as that annihilated by $\widetilde{b}_{\vec{k},s},\widetilde{d}_{-\vec{k},s}$ for all $\vk$, namely

 \be \widetilde{b}_{\vec{k},s}\,\ket{0_O} =0 ~~;~~ \widetilde{d}_{-\vec{k},s}\,\ket{0_O} =0\,.  \label{outvacfer}\ee Pre-multiplying these expressions by $T_f[\theta]$ and inserting $T^{-1}_f[\theta]\,T_f[\theta]=1$,  yields
 \be \underbrace{\Big( T_f[\theta]\,\widetilde{b}_{\vec{k},s}\,T^{-1}[\theta]\Big)}_{b_{\vk}}\,\underbrace{\Big(T[\theta]\,\ket{0_O}\Big)}_{\ket{0_I}} =0 ~~;~~\underbrace{\Big( T[\theta]\,\widetilde{d}_{-\vec{k},s}\,T^{-1}[\theta]\Big)}_{d^\dagger_{-\vk}}\,\underbrace{\Big(T[\theta]\,\ket{0_O}\Big)}_{\ket{0_I}} =0\,.   \label{Toutvacfer}\ee

Applied to the ``out'' vacuum state $\ket{0_O}$ annihilated by $\widetilde{b}_{\vec{k},s},\widetilde{d}_{-\vec{k},s}$ for all $\vk$, we find
\be \ket{0_I}= T_f[\theta]\ket{0_O} = \Pi_{\vk,s}\Big[\cos(\theta_k) - e^{2i\,\varphi_-}\,\sin(\theta_k)\,\widetilde{b}^\dagger_{\vec{k},s}\,\widetilde{d}^\dagger_{-\vec{k},s}\Big]\ket{0_O}\,, \label{Sonvac}\ee where we restored the phases as per equation (\ref{bdphases}). It proves convenient to write this result as
\be \ket{0_I} = \Pi_{\vk,s}\Bigg\{\Big[\cos(\theta_k)\Big]~ \,\sum_{n_{\vk,s}=0}^{1} \Bigg(-e^{2i\varphi_-(k)}\,\tan(\theta_k) \Bigg)^{n_{\vk,s}} \ket{n_{\vk,s};\overline{n}_{-\vk,s}}\Bigg\} \,, \label{vacinfer}\ee where the fermionic ``out'' particle-antiparticle states
\be  \ket{n_{\vk,s};\overline{n}_{-\vk,s}}  = \frac{\Big(\widetilde{b}^\dagger_{\vec{k},s}\Big)^{n_{\vk,s}}}{\sqrt{n_{\vk,s} !}}~
\frac{\Big(\widetilde{d}^\dagger_{-\vec{k},s}\Big)^{n_{\vk,s}}}{\sqrt{n_{\vk,s} !}} \ket{0_O} ~~;~~ n_{\vk,s} = 0, 1\,. \label{noutsapp}\ee Unitarity of the transformation is confirmed by obtaining
\be \langle 0_I|0_I\rangle= \Pi_{\vk,s}\Bigg\{\cos^2(\theta_k)\Big[1+ \tan^2(\theta_k) \Big]  \Bigg\} =1 \,. \label{normaI}\ee Furthermore, we find
\be  \langle 0_I|\widetilde{b}^\dagger_{\vec{k},s}\,\widetilde{b}_{\vec{k},s} |0_I\rangle = \langle 0_I|\widetilde{d}^\dagger_{\vec{k},s}\,\widetilde{d}_{\vec{k},s} |0_I\rangle = \sin^2(\theta_k) = |B_{k,s}|^2 = N_k \,.\label{exvalN}\ee

\section{Summary of adiabatic expansion for fermions:}\label{app:ferad}
In this appendix we provide a brief summary  of the adiabatic expansion for fermions. For more details see ref.\cite{herringfer}   We write generically the spinors as $U$, $V$ with the implicity understanding that during (RD) these are to be identified with the solutions  $\mathcal{U}\,;\,\mathcal{V}$.

 Consider the mode equation (\ref{heqn})
  (we suppress the momentum label and conformal time arguments for ease of
 notation)
 \be h^{''}+ (\omega^2-iM')h =0 \label{modeh}\ee and propose the solution
 \be h(\eta) = e^{-i\int^\eta \Omega(\eta')\,d\eta'} ~~;~~ \Omega= \Omega_R + i\Omega_I \,. \label{solwkb}\ee Introducing this ansatz into the mode equation (\ref{modeh}) yields
 \be \Omega^2 + i\Omega' - \omega^2 + iM' =0 \,, \label{eqnOme}\ee separating the real and imaginary parts yields the coupled system of equations
 \bea && \Omega^2_R - \Omega^2_I - \Omega^{'}_I -\omega^2 = 0 \,\label{realOm}\\
 && 2\Omega_R\Omega_I + (\Omega^{'}_R+M')= 0 ~~ \Rightarrow ~~ \Omega_I = -\frac{(\Omega^{'}_R+M')}{2\Omega_R}\,.  \label{ImagOm}\eea The above equations can be solved in a consistent adiabatic expansion in derivatives of $\omega, M$ with respect to conformal time, we find

  \be \Omega^{(0)}_R = \omega~;~ \Omega^{(0)}_I = 0 ~~;~~  \Omega^{(1)}_R=0 ~;~ \Omega^{(1)}_I = -\frac{(\omega'+M')}{2\omega}~~;~~ \Omega^{(2)}_R = \frac{(\Omega^{(1)}_I)^2 + (\Omega^{(1)}_I)'}{2\omega}~;~\Omega^{(2)}_I = 0 \cdots \,.  \label{lowords}\ee

    In the  representation (\ref{solwkb})  it follows that the spinors can be written compactly as
  \be U_s(\vec{k},\eta) = N\,e^{-i\int^{\eta}\Omega_k(\eta') d\eta'}\,\left( \begin{array}{c}
                                     (\Omega +M )\, \xi_s\\
                                    k  \, s \, \xi_s
                                  \end{array}\right)\,,  \label{Uad}\ee

 \be V_s(-\vec{k},\eta) = N\,e^{i\int^{\eta}\Omega^{*}_k(\eta') d\eta'}\left( \begin{array}{c}
                                      -k  \, s\,\xi_s\\
                                    (\Omega^* +M )  \,   \xi_s
                                  \end{array}\right)\,,  \label{Vad}\ee with $N$ a normalization constant. The orthogonality conditions $U^\dagger_s U_{s'}= 0, V^\dagger_s V_{s'}=0$ for $s\neq s'$ and  $U^\dagger_s \,V_{s'} =0$ for all $s,s'$ are evident.

  Normalizing the spinors $U^\dagger_s U_{s'}= \delta_{s,s'}= V^\dagger_s V_{s'} $  it follows that                                 \be  U_s(\vec{k},\eta) =  \frac{e^{-i\int^{\eta}\Omega_{R}(\eta') d\eta'}}{\Big[\Omega^2_R + \Omega^2_I + \omega^2 +2M\Omega_R \Big]^{1/2}}\,\left( \begin{array}{c}
                                     (\Omega +M )\, \xi_s\\
                                    k  \, s \, \xi_s
                                  \end{array}\right)\,,  \label{Uadfin}\ee

\be V_s(-\vec{k},\eta) =  \frac{e^{i\int^{\eta}\Omega_{R}(\eta') d\eta'}}{\Big[\Omega^2_R + \Omega^2_I + \omega^2 +2M\Omega_R \Big]^{1/2}}\left( \begin{array}{c}
                                      -k  \, s\,\xi_s\\
                                    (\Omega^* +M )  \,   \xi_s
                                  \end{array}\right)\,.  \label{Vadfin}\ee

  To leading (zeroth) adiabatic order with $\Omega_R = \omega_k(\eta), \Omega_I =0$.


\begin{thebibliography}{99}

 \bibitem{bertone} G. Bertone, D. Hooper, J. Silk, , Physics Reports 405,    279 (2005).



\bibitem{neweraDM} G. Bertone, T. M. P. Tait,  Nature 562 (2018) no.7725, 51-56.

    \bibitem{LHCDM} F. Kahlhoefer,  	Int.J.Mod.Phys. A32 (2017) 1730006.

 \bibitem{nowimp1} D. S. Akerib (LUX collaboration),   	Phys. Rev. Lett. 118, 021303 (2017).

 \bibitem{nowimp2} E. Aprile (Xenon Collaboration),   	Phys. Rev. Lett. 121, 111302 (2018).

       \bibitem{parker} L. Parker, Phys. Rev. Lett. \textbf{21}, 562 (1968); Phys. Rev.
       D183, 1057 (1969); Phys. Rev. D3, 346 (1971); J. Phys. A 45, 374023 (2012).

        \bibitem{ford} L. H. Ford, Phys. Rev. D35, 2955 (1987).

         \bibitem{moste1}  A. A. Grib, S. G. Mamayev, V. M. Mostepanenko, Gen.Rel.and Grav. 7, 535 (1976);
    A. A. Griv, B. A. Levitsky, V. M. Mostepanenko, Teor.Mat.Fiz. 19, 59 (1974).

    \bibitem{birrell} N. D. Birrell, P. C. W. Davies, \textit{Quantum fields in curved space time}, (Cambridge Monographs on Mathematical Physics, Cambridge University Press, Cambridge, 1982).



    \bibitem{fullbook} S. A. Fulling, \textit{Aspects of quantum field theory in curved space-time} (Cambridge University Press, Cambridge 1989).

         \bibitem{parkerbook}     L. Parker, D. Toms, \textit{Quantum field theory in curved spacetime: quantized fields and gravity.} (Cambridge Monographs in Mathematical Physics, Cambridge, 2009).

    \bibitem{mukhabook} V. Mukhanov, S. Winitzki, \textit{Introduction to quantum effects in gravity},
    (Cambridge University Press, Cambridge, 2012).



    \bibitem{heavydm1} D. J. H. Chung, E. W. Kolb, A. Riotto, Phys. Rev. D59, 023501 (1999)

        \bibitem{heavydm2} D. J. H. Chung, P. Crotty,  E. W. Kolb, A. Riotto, Phys. Rev. D64, 043503 (2001).

            \bibitem{heavydm3} D. J. H. Chung, E. W. Kolb, A. J. Long,  JHEP 1901,  189 (2019).

      \bibitem{kuzmin} V. Kuzmin, I. Tkachev, Phys. Rev. D59, 123006 (1999);  V. A. Kuzmin and I. I. Tkachev,JETP Lett.68, 271 (1998).

  \bibitem{kuzmin2} V. A. Kuzmin, I. I. Tkachev,   Phys.Rept.320,  199 (1999).

\bibitem{chungfer}  D. J. H. Chung, L. L. Everett, H. Yoo, P. Zhou, Phys. Lett. B712,
  147 (2012).


    \bibitem{ema1} Y. Ema, K. Nakayama, Y. Tang, JHEP 1809, 135 (2018).

\bibitem{ema2}  Y. Ema, R. Jinno, K. Mukaida, K. Nakayama,  Phys. Rev. D 94, 063517 (2016).

  \bibitem{branreh} H. B. Moghaddam, R. Brandenberger,   J. Yokoyama
Phys. Rev. D 95, 063529, (2017).

            \bibitem{vela} J. M. Sanchez-Velazquez, J. A. R. Cembranos, L. J. Garay, JHEP 06, 084 (2020).






\bibitem{hash} S. Hashiba, J. Yokoyama, Phys. Rev. D99, 043008 (2019).

 \bibitem{vilja1}  J. Lankinen, O. Kerppo, I. Vilja,  	Phys. Rev. D 101, 063529 (2020).

 \bibitem{reheat1} For a review: R. Allahverdi, R. Brandenberger, F.-Y. Cyr-Racine,   A. Mazumdar,
Annual Review of Nuclear and Particle Science, 60, 27 (2010).

\bibitem{karam} A. Karam, M. Raidal, E. Tomberg, arXiv:2007.03484.


  \bibitem{reheat} For a review: M. A. Amin,  M. P. Hertzberg, D. I. Kaiser,
J. Karouby,  Int. J. of Mod. Phys. 24, 1530003 (2015).



  \bibitem{vilja2} J. Lankinen, I. Vilja, JCAP  1708, 025 (2017).

\bibitem{herring} N. Herring, D. Boyanovsky, A. Zentner, Phys. Rev. D 101, 083516 (2020).

\bibitem{herringfer}  N. Herring, D. Boyanovsky,   Phys. Rev. D 101, 123522 (2020).

  \bibitem{bernstein} J. Bernstein, \emph{Kinetic theory in the expanding universe}, (Cambridge Monographs on Mathematical Physics, Cambridge University Press, Cambridge, UK, 1988).


\bibitem{gasp} M. Gasperini, M. Giovannini, Class.Quant.Grav.10:L133, (1993).

\bibitem{gasp2}  M. Gasperini, M. Giovannini,  	Phys.Lett. B301, 334 (1993).

\bibitem{gasperini} M. Gasperini, M. Giovannini,  in "String gravity and physics at the Planck energy scale" (World Scientific, Singapore, 1995),  	arXiv:hep-th/9502112.

\bibitem{prokopec} R. Brandenberger, T.Prokopec, V. Mukhanov,  	Phys.Rev. D48,  2443  (1993).

\bibitem{prokobran} R. Brandenberger, V. Mukhanov, T. Prokopec,  	Phys.Rev.Lett. 69, 3606 (1992).


\bibitem{bran1} S. Brahma, O. Alaryani, R. Brandenberger,  	arXiv:2005.09688.

\bibitem{lello}  Louis Lello, Daniel Boyanovsky, Richard Holman,  JHEP04,055 (2014).

\bibitem{boyan} D. Boyanovsky,  Phys. Rev. D 98, 023515 (2018).

\bibitem{beilok} S-Y. Lin, C-H. Chou, B. L. Hu, Phys. Rev.  D 81, 084018 (2010).

\bibitem{martin} E. Martin-Martinez, N. C. Menicucci, Class. Quantum Grav. 29, 224003 (2012).

\bibitem{ball} J. L. Ball, I. Fuentes-Schuller, F P.Schuller, Phys. Lett. A359 (2006) 550

  \bibitem{mann} I.  Fuentes,   R.  B.  Mann,   E.  Martin-Martinez,   S. Moradi, Phys. Rev. D 82, 045030 (2010).

 \bibitem{machado} L. N. Machado, H. A. S. Costa, I. G. da Paz, M. Sampaio, J. B. Araujo, Phys. Rev. D 98, 125009 (2018).


\bibitem{planck2018} Planck collaboration, arXiv: 1807.06211.





\bibitem{gr} I. S. Gradshteyn and I. M. Ryzhik, \textit{Table of Integrals, Series and Products}, (Academic Press, New York, 1980).

 \bibitem{as} M. Abramowitz, I. A. Stegun, \textit{Handbook of Mathematical Functions}, Dover, NY. (1964).

 \bibitem{nist} F. W. Olver, D. W. Lozier, R. F. Boisvert, C. W. Clark, \textit{NIST Handbook of Mathematical Functions}, Cambridge Univ. Press, N.Y. (2010).

     \bibitem{bateman} H. Bateman, \textit{Higher Transcendental Functions, vol. II} (McGraw-Hill, N.Y. 1953).

         \bibitem{magnus} W. Magnus, F. Oberhettinger, R. P. Soni, \textit{Formulas and Theorems
         for the Special Functions of Mathematical Physics.} Springer-Verlag, NY 1966.


 \bibitem{nielsen}  M. A. Nielsen and  I. L. Chuang, \emph{Quantum Computation and Quantum Information}, (Cambridge University Press, Cambridge, UK, 2010).


 \bibitem{bunch} T. S. Bunch, J. Phys. A: Math. Gen. 13, 1297 (1980).

 \bibitem{pf} L. Parker, S. A. Fulling, Phys. Rev. D9, 341 (1974).

 \bibitem{fh} S. A. Fulling, L. Parker, B. L. Hu, Phys. Rev. D10, 3905 (1974).

 \bibitem{hu} B. L. Hu, Phys. Lett. A71, 169 (1979);  B. L. Hu, Phys. Rev. D18, 4460 (1978).

 \bibitem{anderson} P. Anderson, L. Parker, Phys. Rev. D36, 2963 (1987).

 \bibitem{bir} N. D. Birrell, Proc.   R. Soc.   Lond., B361, 513 (1978).

    \bibitem{mottola} S. Habib, C. Molina-Paris, E. Mottola, Phys. Rev. D61, 024010 (1999).



        \bibitem{barnett} S. M. Barnett, P. M. Radmore, \emph{Methods in Theoretical Quantum Optics} (Oxford Science Publications-Clarendon Press, Oxford 1977).

     \bibitem{weinbergbook} S. Weinberg, \emph{Gravitation and Cosmology:
principles and applications of the general
theory of relativity. } (John Wiley , N.Y. 1972).



\bibitem{casta} M. A. Castagnino, L. Chimento, D. D. Harari and C.
Nunez, J. Math. Phys. \textbf{25}, 360 (1984).


 \bibitem{rio} A. del Rio, J. Navarro-Salas, F. Torrenti, Phys. Rev. D90, 084017 (2014).

  \bibitem{ferreiro}  A. Ferreiro, A. del Rio, J. Navarro-Salas, S. Pla, F. Torrenti,
  arXiv:1904.00062.

 \bibitem{barbero} J. Fernando Barbero, A. Ferreiro, J. Navarro-Salas, E. J. S. Villase\~{n}or, Phys. Rev. D98, 025016 (2018).

     \bibitem{ghosh} S. Ghosh, Phys. Rev. D91, 124075 (2015); Phys. Rev. D93, 044032 (2016).

     \bibitem{landete}  A. Landete, J. Navarro-Salas, F. Torrenti,  Phys. Rev. D89, 044030  (2014);
     Phys. Rev. D 88, 061501(R) (2013).




\bibitem{zurek} W. H. Zurek, Phys. Rev. D24, 1516 (1981); Phys. Rev. D26, 1862 (1982);  Rev. Mod. Phys.75, 715 (2003).





 \bibitem{gordon} C. Gordon, D. Wands, B. A. Bassett, R. Maartens, Phys. Rev. D63, 023506 (2000).

 \bibitem{byrnes} C. T. Byrnes, D. Wands, Phys. Rev. D74, 043529 (2006).

 \bibitem{bartolo} N. Bartolo, S. Matarrese, A. Riotto, Phys. Rev. D64, 123504 (2001).

 \bibitem{sena} D. J. H. Chung, E. W. Kolb, A. Riotto, L. Senatore, Phys.Rev. D72, 023511 (2005).

\bibitem{chungiso}  D. J. H. Chung, H. Yoo, P. Zhou, Phys. Rev. D91, 043516 (2015).

\bibitem{lb} D. Lynden-Bell, MNRAS 136, 101 (1967).

\bibitem{tremaine} S. Tremaine, M. H\'enon, D. Lynden-Bell, MNRAS 219, (1986).

\bibitem{decay1}  Nathan Herring, Brian Pardo, Daniel Boyanovsky, Andrew R. Zentner,  Phys. Rev. D 98, 083503 (2018).

\bibitem{decay2}  Daniel Boyanovsky, Nathan Herring,  Phys. Rev. D 100, 023531 (2019).

\bibitem{kandrup} B. L. Hu and Henry E. Kandrup, Phys. Rev. D 35, 1776 (1987).







\end{thebibliography}
\end{document}